\documentclass[preprint,amsmath,superscriptaddress,amssymb]{revtex4}

\usepackage{titlesec}
\usepackage{graphicx,epsfig,epstopdf}
\usepackage{bm}
\usepackage{epstopdf}
\usepackage{amssymb,amsmath,amsfonts,latexsym}
\usepackage{dcolumn}
\usepackage{bm}
\usepackage{hyperref}
\usepackage[italic]{hepnames}
\usepackage[utf8]{inputenc}
\newcommand{\nn}{\nonumber\\}
\usepackage{breqn}             
\makeatletter                  
\let\cat@comma@active\@empty   
\makeatother                   

\newcommand{\bea}{\begin{eqnarray}}
\newcommand{\ena}{\end{eqnarray}}
\newcommand{\be}{\begin{equation}}
\newcommand{\en}{\end{equation}}
\newcommand{\Tr}{\mbox{\rm{tr}}}

\begin{document}
\title{Semileptonic $D_{(s)}$-meson decays in the light of recent data}

\author{N.~R.~Soni}
\email{nrsoni-apphy@msubaroda.ac.in}
\affiliation{Applied Physics Department, Faculty of Technology and Engineering, \\ The Maharaja Sayajirao University of Baroda, Vadodara 390001, Gujarat, India}

\author{M.~A.~Ivanov}
\email{ivanovm@theor.jinr.ru}
\affiliation{Bogoliubov Laboratory of Theoretical Physics, \\
Joint Institute for Nuclear Research, 141980 Dubna, Russia}

\author{J.~G.~K\"{o}rner}
\email{jukoerne@uni-mainz.de}
\affiliation{PRISMA Cluster of Excellence, Institut f\"{u}r Physik, \\
Johannes Gutenberg-Universit\"{a}t, 
D-55099 Mainz, Germany}

\author{J.~N.~Pandya}
\email{jnpandya-apphy@msubaroda.ac.in}
\affiliation{Applied Physics Department, Faculty of Technology and Engineering, \\ The Maharaja Sayajirao University of Baroda, Vadodara 390001, Gujarat, India}

\author{P.~Santorelli}
\email{Pietro.Santorelli@na.infn.it}
\affiliation{Dipartimento di Fisica ``E.~Pancini'', Universit\`{a} di Napoli Federico II,\\ Complesso Universitario di Monte S. Angelo, Via Cintia, Edificio 6, 80126 Napoli, Italy}
\affiliation{Istituto Nazionale di Fisica Nucleare, Sezione di Napoli, 80126 Napoli, Italy}

\author{C.~T.~Tran}
\email{tranchienthang1347@gmail.com}
\thanks{corresponding author}
\affiliation{Institute of Research and Development, Duy Tan University, 550000 Da Nang, Vietnam}
\affiliation{Dipartimento di Fisica ``E.~Pancini'', Universit\`{a} di Napoli Federico II,\\ Complesso Universitario di Monte S. Angelo, Via Cintia, Edificio 6, 80126 Napoli, Italy}

\date{\today}

\begin{abstract}
Inspired by recent improved measurements of charm semileptonic decays at BESIII, we study a large set of 
$D (D_s)$-meson semileptonic decays where
the hadron in the final state is one of $D^0$, $\rho$, $\omega$, $\eta^{(\prime)}$ in the case of $D^+$ decays, and $D^0$, $\phi$, $K^0$, $K^\ast(892)^0$, $\eta^{(\prime)}$ in the case of $D^+_s$ decays. The required hadronic form factors are computed in the full kinematical range of momentum transfer by employing the covariant confined quark model developed by us. A detailed comparison of the form factors with those from other approaches is provided. We calculate the decay branching fractions and their ratios, which show good agreement with available experimental data. We also give predictions for the forward-backward asymmetry and the longitudinal and transverse polarizations of the charged lepton in the final state.

 \end{abstract}

\maketitle


\section{Introduction}
\label{sec:introduction}

Semileptonic $D(D_s)$-meson decays provide a good platform to study both the weak and strong interactions in the charm sector (for a review, see e.g., Ref.~\cite{Richman:1995wm}). Measurements of their decay rates allow a direct determination of the Cabibbo-Kobayashi-Maskawa (CKM) matrix elements $|V_{cs}|$ and $|V_{cd}|$. In particular, the average of the measurements of {\it BABAR}~\cite{Lees:2014ihu, Aubert:2007wg}, Belle~\cite{Widhalm:2006wz}, BESIII~\cite{Ablikim:2015ixa}, and CLEO~\cite{Besson:2009uv} of the decays $D\to\pi(K)\ell\nu$ was used to extract the elements $|V_{cd(s)}|$, as recently reported by the Particle Data Group (PDG)~\cite{Tanabashi:2018oca}. Such extraction of the CKM matrix elements from experiments requires theoretical knowledge of the hadronic form factors which take into account the nonperturbative quantum chromodynamics (QCD) effects. 

The elements $|V_{cs}|$ and $|V_{cd}|$ can also be determined indirectly by using the unitarity constraint on the CKM matrix. This method was very useful in the past when the direct measurements still suffered from large uncertainties, both experimental and theoretical. Once these matrix elements are determined, whether directly or indirectly, one can in reverse study the strong interaction effects in various charm semileptonic channels to reveal the decay dynamics. One can also test the predictions of different theoretical approaches, such as the form factors and the branching fractions. In this manner, the study of semileptonic charm decays can indirectly contribute to a more precise determination of other CKM matrix elements such as $|V_{ub}|$, in the sense that constraints provided by charm decays can improve the theoretical inputs needed for extracting $|V_{ub}|$ from exclusive charmless $B$ semileptonic decays.

Recent progresses in experimental facilities and theoretical studies have made more and more stringent tests of the standard model (SM) available in the charm sector and have opened a new window through which to look for possible new physics effects beyond the SM. These tests include the CKM matrix unitarity, CP violations, isospin symmetry, and lepton flavor universality (LFU). Notably, the BESIII collaboration has reported recently measurements of many semimuonic charm decays~\cite{Ablikim:2016sqt, Ablikim:2017omq, Ablikim:2018frk}, some for the first time and some with much improved precision. This paves the way to the search for signals of LFU violations in these channels. In addition, the study of the decays $D_s \to \eta^{(\prime)} \ell^+\nu_\ell$ provides information about the $\eta-\eta^\prime$ mixing angle and helps probe the interesting $\eta-\eta^\prime$-glueball mixing~\cite{Anisovich:1997dz, DiDonato:2011kr}.

From the theoretical point of view, the calculation of hadronic form factors plays a crucial role in the study of charm semileptonic decays. This calculation is carried out by nonperturbative methods including lattice QCD (LQCD)~\cite{Donald:2013pea, Bali:2014pva, Aoki:2016frl}, QCD sum 
rules~\cite{Ball:1993tp, Colangelo:2001cv, Du:2003ja}, light-cone sum rules (LCSR)~\cite{Khodjamirian:2000ds, Wu:2006rd, Azizi:2010zj, Offen:2013nma,Meissner:2013hya, Duplancic:2015zna, Fu:2018yin}, and phenomenological quark models. Regarding the quark models used in studies of semileptonic $D$ decays, one can mention the Isgur-Scora-Grinstein-Wise (ISGW) model~\cite{Isgur:1988gb} and its updated version ISGW2~\cite{Scora:1995ty}, the constituent quark model (CQM)~\cite{Melikhov:2000yu}, the relativistic quark model based on the quasipotential approach~\cite{Faustov:1996xe}, the chiral quark model~\cite{Palmer:2013yia}, the light-front quark model (LFQM)~\cite{Wei:2009nc, Verma:2011yw, Cheng:2017pcq}, and the model based on the combination of heavy meson and chiral symmetries (HM$\chi$T)~\cite{Fajfer:2004mv, Fajfer:2005ug}. 
Several semileptonic decay channels of the $D_{(s)}$ mesons were also studied in the large energy effective theory~\cite{Charles:1998dr}, chiral perturbation theory~\cite{Bijnens:2010jg}, the so-called chiral unitary approach ($\chi$UA)~\cite{Sekihara:2015iha}, and a new approach assuming pure heavy quark symmetry~\cite{Dai:2018vzz}. Recently, a simple expression for $D\to K$ semileptonic form factors was studied in Ref.~\cite{Pham:2018bxs}. We also mention here early attempts to account for flavor symmetry breaking in pseudoscalar meson decay constants by the authors of Ref.~\cite{Khlopov}. It is worth noting that each method has only a limited range of applicability, and their combination will give a better picture of the underlined physics~\cite{Melikhov:2000yu}. 

In this paper, we compute the form factors of the semileptonic $D(D_s)$ decays in the framework of the covariant confined quark model (CCQM)~\cite{Efimov, Branz:2009cd, Ivanov:2011aa, Gutsche:2012ze}. To be more specific, we study the decays $D^+\to (D^0, \rho^0, \omega, \eta,\eta^\prime)\ell^+\nu_\ell$, $D_s^+\to (D^0, \phi, K^0, K^\ast(892)^0, \eta, \eta^\prime)\ell^+\nu_\ell$, and $D^0\to \rho^-\ell^+\nu_\ell$. This paper follows our previous study~\cite{Soni:2017eug} in which some of us have considered the decays $D\to K^{(\ast)}\ell^+\nu_\ell$ and $D\to \pi\ell^+\nu_\ell$ in great detail. 
Our aim is to provide a systematic and independent study of $D_{(s)}$ semileptonic channels in the same theoretical framework. This will shed more light on the theoretical study of the charm decays, especially on the shape of the corresponding form factors, since the CCQM  predicts the form factors in the whole physical range of momentum transfer without using any extrapolations. Besides, many of the studies mentioned in the previous paragraph were done about a decade ago, with the main focus on the branching fraction. In light of recent data, more up-to-date predictions are necessary, not only for the branching fraction but also for other physical observables such as the forward-backward asymmetry and the lepton polarization. Finally, such a systematic study is necessary to test our model's predictions and to better estimate its theoretical error.

The rest of the paper is organized as follows. In Sec.~\ref{sec:formalism}, we briefly provide the definitions of the semileptonic matrix element and hadronic form factors. Then we give the decay distribution in terms of the helicity amplitudes. In Sec~\ref{sec:formfact}, we introduce the essential ingredients of the covariant confined quark model and describe in some detail the calculation of the form factors in our approach. Numerical results for the form factors, the decay branching fractions, and other physical observables are presented in Sec.~\ref{sec:results}. We compare our findings with other theoretical approaches as well as experimental data including recent LQCD calculations and BESIII data. Finally, the conclusion is given in Sec.~\ref{sec:conclusion}.
\section{Matrix element and decay distribution}
\label{sec:formalism}

Within the SM, the matrix element for semileptonic decays of the $D_{(s)}$ meson to a pseudoscalar ($P$) or a vector ($V$) meson in the final state is written as
\be
\mathcal M(D_{(s)} \to (P,V) \ell^+ \nu_{\ell}) = \frac{G_F}{\sqrt{2}} V_{cq} \langle (P,V) | \bar{q} O^\mu c | D_{(s)} \rangle  
[\ell^+ O_\mu \nu_{\ell}],
\en
where $O^\mu=\gamma^\mu(1-\gamma_5)$, and $q=d, s$.
The hadronic part in the matrix element is parametrized by the invariant form factors which depend on the momentum transfer squared $q^2$ between the two mesons as follows:
\bea
\label{eq:formfac}
\langle P(p_2)
|\bar{q} O^\mu c
| D_{(s)}(p_1) \rangle
&=& F_+(q^2) P^\mu + F_-(q^2) q^\mu,\nn
\langle V(p_2,\epsilon_2)
|\bar{q} O^\mu c
| D_{(s)}(p_1) \rangle
&=& \frac{\epsilon^{\dagger}_{2\alpha}}{M_1+M_2}
\Big[-g^{\mu\alpha}PqA_0(q^2) + P^{\mu}P^{\alpha}A_+(q^2)\\
&&+ q^{\mu}P^\alpha A_-(q^2) 
+ i\varepsilon^{\mu\alpha Pq} V(q^2)\Big],
\nonumber
\ena
where $P = p_1 + p_2$, $q=p_1-p_2$, and $\epsilon_2$ is the polarization vector of the vector meson $V$, so that $\epsilon_2^\dagger\cdot p_2=0$. The mesons are on shell: $p_1^2=m^2_{D_{(s)}}=M_1^2$, $p_2^2=m_{P,V}^2=M_2^2$.

For later comparison of the form factors with other studies, we relate our form factors defined in Eq.~(\ref{eq:formfac}) to the well-known Bauer-Stech-Wirbel (BSW) form factors~\cite{Wirbel:1985ji}, namely, $F_{+,0}$ for $D_{(s)} \to P$ and $A_{0,1,2}$ and $V$ for $D_{(s)} \to V$.  Note that in Ref.~\cite{Wirbel:1985ji} the notation $F_1$ was used instead of $F_+$. The relations read
\bea
\widetilde{A}_2 &=& A_+, \qquad \widetilde{V} = V,\qquad \widetilde{F}_+ = F_+, \nn
\widetilde{A}_1 &=& \frac{M_1-M_2}{M_1+M_2} A_0,\qquad
\widetilde{F}_0 = F_+ + \frac{q^2}{M_1^2-M_2^2} F_-, \\
\widetilde{A}_0 &=& \frac{M_1-M_2}{2M_2}\Big(A_0-A_+ - \frac{q^2}{M_1^2-M_2^2} A_- \Big).\nonumber
\label{eq:ff-rel}
\ena
Here,  the BSW form factors are denoted with a tilde to distinguish from our form factors. However, for simplicity, we will omit the tilde in what follows. In all comparisons of the form factors to appear below, we use the BSW ones.

Once the form factors are known, one can easily calculate the semileptonic decay rates. However, it is more convenient to write down the differential decay width in terms of the so-called helicity amplitudes which are combinations of the form factors. This is known as the helicity technique, first described in Ref.~\cite{Korner-Schuler} and further discussed in our recent papers~\cite{Gutsche:2015mxa, Ivanov:2015tru}. One has
\bea
\frac{d\Gamma(D_{(s)} \to (P,V) \ell^+ \nu_\ell)}{dq^2} &=& \frac{G_F^2 |V_{cq}|^2 |{\bf p_2}| q^2}{96 \pi^3M_1^2}  \Big(1-\frac{m_\ell^2}{q^2} \Big)^2 \nn
&&\times \Big[\Big(1+\frac{m_\ell^2}{2q^2}\Big) (|H_+|^2 + |H_-|^2 + |H_0|^2) + \frac{3m_\ell^2}{2q^2} |H_t|^2\Big],
\label{eq:decay_width}
\ena
where $|{\bf p_2}| = \lambda^{1/2} (M_1^2, M_2^2, q^2)/2 M_1$ is the momentum of the daughter meson in the rest frame of the parent meson. Here, the helicity amplitudes for the decays $D_{(s)} \to V \ell^+ \nu_\ell$ are defined as
\bea
H_{\pm} &=& \frac{1}{M_1 + M_2} \left(-Pq A_0 \pm 2 M_1 |{\bf p_2}| V\right),\nn
H_{0} &=& \frac{1}{M_1 + M_2} \frac{1}{2 M_2\sqrt{q^2}} \left[-Pq (M_1^2 - M_2^2 - q^2) A_0 + 4 M_1^2 |{\bf p_2}|^2 A_+\right],\\
H_{t} &=& \frac{1}{M_1 + M_2} \frac{M_1 |\bf{p_2}|}{M_2\sqrt{q^2}} \left[Pq (-A_0 + A_+) + q^2 A_-\right].\nonumber
\ena
In the case of the decays $D_{(s)} \to P \ell^+ \nu_\ell$ one has
\be
H_\pm = 0, \qquad 
H_0 = \frac{2 M_1 |\bf{p_2}|}{\sqrt{q^2}} F_+, \qquad 
H_t = \frac{1}{\sqrt{q^2}} (Pq F_+ + q^2 F_-).
\en

In order to study the lepton-mass effects, one can define several physical observables such as the forward-backward asymmetry $\mathcal{A}_{FB}^\ell(q^2)$ and the longitudinal $P_L^\ell(q^2)$ and transverse $P_T^\ell(q^2)$ polarization of the charged lepton in the final state. This requires the angular decay distribution, which was described elsewhere~\cite{Ivanov:2015tru}. In short, one can write down these observables in terms of the helicity amplitudes as follows:
\bea
\mathcal{A}_{FB}^\ell(q^2) &=&
-\frac{3}{4}
\frac{|H_+|^2-|H_-|^2+4\delta_\ell H_0 H_t}{(1+\delta_\ell)\sum |H_n|^2+3\delta_\ell |H_t|^2},
\label{eq:FB}
\\
P_L^\ell(q^2) &=& 
-\frac{(1-\delta_\ell)\sum |H_n|^2-3\delta_\ell |H_t|^2}{(1+\delta_\ell)\sum |H_n|^2+3\delta_\ell |H_t|^2},
\label{eq:PL}\\
P_T^\ell(q^2) &=& 
-\frac{3\pi}{4\sqrt{2}}\frac{\sqrt{\delta_\ell}(|H_+|^2-|H_-|^2-2H_0 H_t)}{(1+\delta_\ell)\sum |H_n|^2+3\delta_\ell |H_t|^2},
\label{eq:PT}
\ena
where $\delta_\ell=m_\ell^2/2q^2$ is the helicity-flip factor, and the index $n$ runs through $(+, -,0)$. The average of these observables over the $q^2$ range is better suited for experimental measurements with low statistics. To calculate the average one has to multiply the numerator and denominator of e.g. Eq.~(\ref{eq:FB}) by the phase-space factor $C(q^2)=|{\bf p_2}|(q^2-m_\ell^2)^2/q^2$ and integrate them separately. These observables are  sensitive to contributions of physics beyond the SM and can be used to test LFU violations~\cite{Bifani:2018zmi, Ivanov:2017mrj, Hu:2018veh, Asadi:2018sym, Rajeev:2018txm, Feruglio:2018fxo, Alonso:2018vwa}. 

\section{Form factors in the covariant confined quark model}
\label{sec:formfact}
In this study, the semileptonic form factors are calculated in the framework of the CCQM~\cite{Efimov, Branz:2009cd}. The CCQM is an effective quantum field approach to the calculation of hadronic transitions. The model is built on the assumption that hadrons interact via constituent quark exchange only. This is realized by adopting a relativistic invariant Lagrangian that describes the coupling of a hadron to its constituent quarks. This approach can be used to treat not only mesons~\cite{Ivanov:1999ic, Faessler:2002ut, Ivanov:2006ni, Ivanov:2015woa, Dubnicka:2018gqg}, but also baryons~\cite{Gutsche:2013pp, Gutsche:2017hux, Gutsche:2018nks}, tetraquarks~\cite{Dubnicka:2011mm, Goerke:2016hxf, Goerke:2017svb}, and other multiquark states~\cite{Gutsche:2017twh} in a consistent way. For a detailed description of the model and the calculation techniques we refer the reader to the references mentioned above. We list below only several key features of the CCQM for completeness.

For the simplest hadronic system, i.e. a meson $M$, the interaction Lagrangian is given by
\begin{dmath}\label{eq:int_lagrange}
\mathcal{L}_{\rm int}  =  g_M M(x) \!\! \int \!\! dx_1 dx_2 F_M(x;x_1,x_2) \bar{q}_2(x_2) \Gamma_M q_1(x_1) +  {\rm H.c.},
\end{dmath}
where $g_M$ is the quark-meson coupling and $\Gamma_M$ the Dirac matrix. For a pseudoscalar (vector) meson $\Gamma_M = \gamma_5$ ($\Gamma_M = \gamma_\mu$). The vertex function $F_M(x,x_1,x_2)$ effectively describes the quark distribution in the meson and is given by
\begin{equation}\label{eq:vertex_function}
F_M(x,x_1,x_2) = \delta \Big(x - \sum_{i=1}^2 w_i x_i \Big)\cdot\Phi_M \big((x_1 - x_2)^2\big),
\end{equation}
where $w_{q_i} = m_{q_i}/ (m_{q_1} + m_{q_2})$ such that $w_1 + w_2 = 1$. The function $\Phi_M$ depends on the effective size of the meson. In order to avoid ultraviolet divergences in the quark loop integrals, it is required that the Fourier transform of $\Phi_M$ has an appropriate falloff behavior in the Euclidean region. Since the final results are not sensitive to the specific form of $\Phi_M$, for simplicity, we choose a Gaussian form as follows:
\begin{equation} \label{eq:gaussian}
\widetilde{\Phi}_M(-p^2) = \!\! \int \!\! dx e^{ipx}\Phi_M(x^2)=e^{p^2/\Lambda_M^2},
\end{equation}
where the parameter $\Lambda_M$ characterizes the finite size of the meson. 

The coupling strength $g_M$ is determined by the compositeness condition $Z_M=0$~\cite{Z=0}, where $Z_M$ is the wave function renormalization constant of the meson. This condition ensures the absence of any bare quark state in the physical mesonic state and, therefore, helps avoid double counting and provides an effective description of a bound state.

In order to calculate the form factors, one first writes down the matrix element of the hadronic transition. In the CCQM, the hadronic matrix element is described by the one-loop Feynman diagram depicted in Fig.~\ref{fig:mass} and is constructed from the convolution of quark propagators and vertex functions as follows:
\begin{figure}[t]
\includegraphics[width=0.65\textwidth]{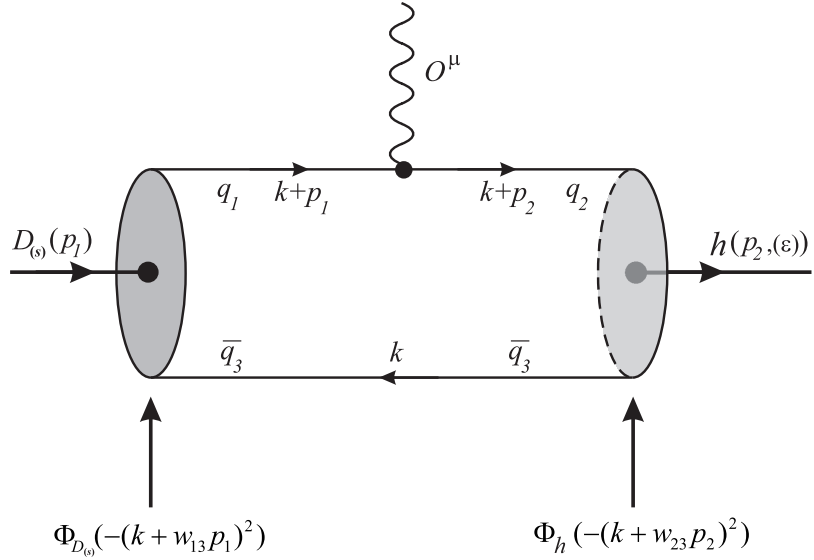}
\vspace*{-5mm}
\caption{Quark model diagram for the $D_{(s)}$-meson semileptonic decay.}
\label{fig:mass}
\end{figure}
\bea
\langle P(p_2)
|\bar{q} O^\mu c
| D_{(s)}(p_1) \rangle
&=&
N_c\, g_{D_{(s)}} g_P \!\! \int \!\!\frac{d^4k}{ (2\pi)^4 i} 
\widetilde\Phi_{D_{(s)}}\!\! \left(-(k+w_{13} p_1)^2\right)\,
\widetilde\Phi_P\left(-(k+w_{23} p_2)^2\right)
\nn
&&\times
\Tr \big[
O^{\mu} S_1(k+p_1) \gamma^5 S_3(k) \gamma^5 S_2(k+p_2) 
\big],
\\
\label{eq:PP'}
\langle V(p_2,\epsilon_2)
|\bar{q} O^\mu c
| D_{(s)}(p_1) \rangle
&=&
N_c\, g_{D_{(s)}} g_V \!\! \int\!\! \frac{d^4k}{ (2\pi)^4 i}\, 
\widetilde\Phi_{D_{(s)}} \!\! \left(-(k+w_{13} p_1)^2\right)\,
\widetilde\Phi_V\left(-(k+w_{23}p_2)^2\right)
\nn
&&\times
\Tr \big[ 
O^{\mu} S_1(k+p_1)\gamma^5 S_3(k) \not\!\epsilon_2^{\,\,\dagger} S_2(k+p_2) \big],
\label{eq:PV}
\ena
where $N_c=3$ is the number of colors, $w_{ij}=m_{q_j}/(m_{q_i}+m_{q_j})$, and $S_{1,2}$ are quark propagators, for which we use the Fock-Schwinger representation
\be
S_i(k) = (m_{q_i}+\not\! k)
\int\limits_0^\infty\! d\alpha_i \exp[-\alpha_i(m_{q_i}^2-k^2)].
\en
It should be noted that all loop integrations  are carried out in Euclidean space.

Using various techniques described in our previous papers, a form factor $F$ can be finally written in the form of a threefold integral
\be
F   =N_c\, g_{D_{(s)}} g_{(P,V)} \!\! \int\limits_0^{1/\lambda^2}\!\! dt\, t
\!\! \int\limits_0^1\!\! d\alpha_1
\!\! \int\limits_0^1\!\! d\alpha_2  \,
\delta\Big(1 -  \alpha_1-\alpha_2 \Big) 
f(t\alpha_1,t\alpha_2),
\label{eq:3fold}
\en
where $f(t\alpha_1,t\alpha_2)$ is the resulting integrand corresponding to the form factor $F$, and $\lambda$ is the so-called infrared cutoff parameter, which is introduced to avoid the appearance of the branching point corresponding to the creation of free quarks and taken to be universal for all physical processes.

The model parameters, namely, the meson size parameters, the constituent quark masses, and the infrared cutoff parameter are determined by fitting the radiative and leptonic decay constants to experimental data or LQCD calculations. The model parameters required for the calculation in this paper are listed in Tables~\ref{tab:size_parameter} and~\ref{tab:quark_mass}.
Other parameters such as the mass and lifetime of mesons and leptons, the CKM matrix elements, and physical constants are taken from the recent report of the PDG~\cite{Tanabashi:2018oca}. In particular, we adopt the following values for the CKM matrix elements: $|V_{cd}|=0.218$ and $|V_{cs}|=0.997$.
\begin{table*}
\caption{Meson size parameters in GeV.}\label{tab:size_parameter}
\renewcommand{\arraystretch}{0.7}
\begin{ruledtabular}
\begin{tabular}{ccccccccccc}
$\Lambda_{D}$ & $\Lambda_{D_s}$ & $\Lambda_{K}$ & $\Lambda_{K^{*}}$ & $\Lambda_\phi$ & $\Lambda_\rho$ & $\Lambda_\omega$ & $\Lambda_{\eta}^{q\bar{q}}$ & $\Lambda_{\eta}^{s\bar{s}}$  & $\Lambda_{\eta'}^{q\bar{q}}$ & $\Lambda_{\eta'}^{s\bar{s}}$\\
\hline
1.600 & 1.750 & 1.014 & 0.805 & 0.880 & 0.610 & 0.488 & 0.881 & 1.973 & 0.257 & 2.797
\end{tabular}
\end{ruledtabular}
\end{table*}
\begin{table}
\caption{Quark masses and infrared cutoff parameter in GeV.}\label{tab:quark_mass}
\renewcommand{\arraystretch}{0.7}
\begin{ruledtabular}
\begin{tabular}{ccccc}
$m_{u/d}$ & $m_s$ &  $m_c$ &  $m_b$ & $\lambda$ \\
\hline
0.241 & 0.428 & 1.672 & 5.05 & 0.181
\end{tabular}
\end{ruledtabular}
\end{table}
\begin{table}[!htb]
\caption{Parameters of the double-pole parametrization 
Eq.~(\ref{eq:double_pole}) for the form factors.} \label{tab:form_factors}
\renewcommand{\arraystretch}{0.7}
\begin{ruledtabular}
\begin{tabular}{lccrlrcc}
$F$ & $F(0)$ & $a$ & $b$ & $F$ & $F(0)$ & $a$ & $b$\\
\hline
$A_+^{D\to\rho}$ 		& 0.57 	& 0.96 	& 0.15		& $A_-^{D\to\rho}$ 		& $-0.74$ 	& 1.11 	& 0.22\\
$A_0^{D\to\rho}$ 		& 1.47 	& 0.47 	& $-0.10$ 	& $V^{D\to\rho}$ 			& 0.76 		& 1.13 	& 0.23\\
$A_+^{D\to\omega}$ 	& 0.55 	& 1.01 	& 0.17 		& $A_-^{D\to\omega}$	& $-0.69$ 	& 1.17 	& 0.26\\
$A_0^{D\to\omega}$ 	& 1.41 	& 0.53 	& $-0.10$ 	& $V^{D\to\omega}$ 		& 0.72 		&1.19 	& 0.27\\
$A_+^{D_s\to\phi}$	& 0.67	& 1.06	& 0.17		& $A_-^{D_s\to\phi}$		& $-0.95$	& 1.20	& 0.26\\
$A_0^{D_s\to\phi}$	& 2.13	& 0.59	& $-0.12$	& $V^{D_s\to\phi}$			& 0.91		& 1.20	& 0.25\\
$A_+^{D_s\to K^*}$	& 0.57	& 1.13 	& 0.21		& $A_-^{D_s\to K^*}$		& $-0.82$	& 1.32	& 0.34\\
$A_0^{D_s\to K^*}$	& 1.53	& 0.61	& $-0.11$	& $V^{D_s\to K^*}$		& 0.80		& 1.32	& 0.33\\
$F_+^{D \to \eta}$		& 0.67 	& 0.93	& 0.12		& $F_-^{D \to \eta}$		& $-0.37$	& 1.02	& 0.18\\
$F_+^{D \to \eta'}$		& 0.76	& 1.23	& 0.23		& $F_-^{D \to \eta'}$		& $-0.064$& 2.29	& 1.71\\
$F_+^{D \to D^0}$		& 0.91	& 5.88	& 4.40		& $F_-^{D \to D^0}$		& $-0.026$& 6.32	& 8.37\\
$F_+^{D_s \to \eta}$	& 0.78	& 0.69	& 0.002		& $F_-^{D_s \to \eta}$	& $-0.42$	& 0.74	& 0.008\\
$F_+^{D_s \to \eta'}$	& 0.73	& 0.88	& 0.018		& $F_-^{D_s \to \eta'}$	& $-0.28$	& 0.92	& 0.009\\
$F_+^{D_s\to K}$		& 0.60 	& 1.05	& 0.18		& $F_-^{D_s\to K}$ 		& $-0.38$	& 1.14	& 0.24\\
$F_+^{D_s \to D^0}$	& 0.92	& 5.08	& 2.25		& $F_-^{D_s \to D^0}$ 	& $-0.34$	& 6.79	& 8.91
\end{tabular}
\end{ruledtabular}
\end{table}

Once the model parameters are fixed, the form factors are obtained by calculating the threefold integral in Eq.~(\ref{eq:3fold}). This is done by using \textsc{mathematica} as well as \textsc{fortran} code. In the CCQM, the form factors are calculable in the entire range of momentum transfer. The calculated form factors are very well represented by the double-pole parametrization
\begin{equation}
\label{eq:double_pole}
F(q^2) = \frac{F(0)}{1-a \hat{s} + b \hat{s}^2},\qquad \hat{s}=\frac{q^2}{m_{D_{(s)}}^2}.
\end{equation}
Our results for the parameters $F(0)$, $a$, and $b$ appearing in the parametrization Eq.~(\ref{eq:double_pole}) are given in Table~\ref{tab:form_factors}.

It is worth noting here that in the calculation of the $D_{(s)}\to\eta^{(\prime)}$ form factors one has to take into account the mixing of the light and the $s$-quark components. By assuming $m_u=m_d\equiv m_q$, the quark content can be written as
\bea
\left(\begin{array}{c} \eta\\
\eta^\prime \end{array}\right) =-\left(\begin{array}{cc} \sin\delta & \cos\delta\\
-\cos\delta & \sin\delta \end{array}\right)
\left(\begin{array}{c} q\bar{q}\\
s\bar{s} \end{array}\right),\qquad
q\bar{q}\equiv \frac{u\bar{u}+d\bar{d}}{\sqrt{2}}.
\ena
The angle $\delta$ is defined by $\delta=\theta_P-\theta_I$, where $\theta_I=\arctan(1/\sqrt{2})$ is the ideal mixing angle. We adopt the value $\theta_P=-15.4^\circ$ from Ref.~\cite{Feldmann:1998vh}.

\section{Results and Discussion}
\label{sec:results}
\subsection{Form factors}
\label{subsec:FF}
In this subsection, we compare our form factors with those from other theoretical approaches and from experimental measurements.
For convenience, we relate all form factors from different studies to the BSW form factors, as mentioned in Sec.~\ref{sec:formalism}. In the SM, the hadronic matrix element between two mesons is parametrized by two form factors ($F_+$ and $F_0$) for the $P\to P^\prime$ transition and four form factors ($A_{0,1,2}$ and $V$) for the $P\to V$ one. However, in semileptonic decays of $D$ and $D_s$ mesons, the form factors $F_0$ and $A_0$ are less interesting because their contributions to the decay rate vanish in the zero lepton-mass limit (the tau mode is kinematically forbidden). Therefore, we focus more on the form factors $F_+$, $A_1$, $A_2$, and $V$. We note that the uncertainties of our form factors mainly come from the errors of the model parameters. These parameters are determined from a
least-squares fit to available experimental data and some lattice calculations.
We have observed that the errors of the fitted parameters are within 10$\%$. We then
calculated the propagation of these errors on the form factors and found the uncertainties on the form factors to be of order 20$\%$ at small $q^2$ and 30$\%$ at high $q^2$. At maximum recoil $q^2=0$, the form factor uncertainties are of order 15$\%$.

We start with the $D_{(s)}\to P$ transition form factor $F_+(q^2)$. In Table~\ref{tab:form_factors_comparison_DP}, we compare the maximum-recoil values $F_+(q^2=0)$ with other theoretical approaches. It is observed that our results are in good agreement with other quark models, especially with the CQM~\cite{Melikhov:2000yu} and the LFQM~\cite{Verma:2011yw}. Besides, quark model predictions for $F_+(0)$ of the $D_{(s)}\to \eta^{(\prime)}$ channels are in general higher than those obtained by LCSR~\cite{Offen:2013nma, Duplancic:2015zna} and LQCD~\cite{Bali:2014pva}. This suggests that more studies of these form factors are needed. For example, a better LQCD calculation of $F_+(0)$ is expected. Note that the authors of Ref.~\cite{Bali:2014pva} considered their LQCD calculation as a pilot study rather than a conclusive one.
\begin{table*}[!htb]
\caption{Comparison of $F_+(0)$ for $D_{(s)} \to P$ transitions.}\label{tab:form_factors_comparison_DP}
\renewcommand{\arraystretch}{0.7}
\begin{ruledtabular}
\begin{tabular}{lccccc}
& $D \to \eta$ & $D \to \eta'$ & $D_s \to \eta$ & $D_s \to \eta'$ & $D_s \to K^0$\\
 \hline
Present& $0.67\pm 0.10$ & $0.76\pm 0.11$ & $0.78\pm 0.12$ & $0.73\pm 0.11$ & $0.60\pm 0.09$\\
CQM~\cite{Melikhov:2000yu} & $\dots$ & $\dots$ & 0.78 & 0.78 & 0.72\\
LFQM~\cite{Verma:2011yw} & 0.71 & $\dots$ & 0.76 & $\dots$ & 0.66\\
LQCD$_{M_\pi = 470\, {\rm MeV}} $\cite{Bali:2014pva} & $\dots$ & $\dots$ & $0.564 (11)$ & $0.437(18)$ & $\dots$\\
LQCD$_{M_\pi = 370 \, {\rm MeV}} $\cite{Bali:2014pva} & $\dots$ & $\dots$ & $0.542(13)$ & $0.404(25)$ & $\dots$\\
LCSR~\cite{Offen:2013nma} & $0.552\pm 0.051$ & $0.458\pm 0.105$ & $0.432\pm 0.033$ & $0.520\pm 0.080$ & $\dots$\\
LCSR~\cite{Duplancic:2015zna} & $0.429^{+0.165}_{-0.141}$ & $0.292^{+0.113}_{-0.104}$ & $0.495^{+0.030}_{-0.029}$ & $0.558^{+0.047}_{-0.045}$ & $\dots$
\end{tabular}
\end{ruledtabular}
\end{table*}
\begin{table}[!htbp]
\caption{Ratios of the $D_{(s)}\to V$ transition form factors at maximum recoil.}\label{tab:form_factors_comparison_DV}
\renewcommand{\arraystretch}{0.7}
\begin{ruledtabular}
\begin{tabular}{lccccccc}
Channel & Ratio & Present & PDG~\cite{Tanabashi:2018oca}   & LQCD~\cite{Donald:2013pea} & CQM~\cite{Melikhov:2000yu} & LFQM~\cite{Verma:2011yw} & HM$\chi$T~\cite{Fajfer:2005ug}\\
\hline
$D \to \rho$			& $r_2$		& $0.93\pm 0.19$ 	& $0.83 \pm 0.12$	& $\dots$ & 0.83 & 0.78 & 0.51\\
 							& $r_V$ 	& $1.26\pm 0.25$ 	& $1.48 \pm 0.16$	 & $\dots$ & 1.53 & 1.47 & 1.72\\
$D^+ \to \omega$	& $r_2$ 	& $0.95\pm 0.19$ 	& $1.06 \pm 0.16$	 & $\dots$ & $\dots$ & 0.84 & 0.51\\
 							& $r_V$ 	& $1.24\pm 0.25$ 	& $1.24 \pm 0.11$	 & $\dots$ & $\dots$ & 1.47 & 1.72\\
$D_s^+ \to \phi$	& $r_2$ 	& $0.99\pm 0.20$	& $0.84 \pm 0.11$	& $0.74(12)$ & 0.73 & 0.86 & 0.52\\
							& $r_V$	& $1.34\pm 0.27$	& $1.80 \pm 0.08$	& $1.72(21)$ & 1.72 & 1.42 & 1.80\\
$D_s^+ \to K^{*0}$		& $r_2$ 	& $0.99\pm 0.20$	& $\dots$ & $\dots$ & 0.74 & 0.82 & 0.55\\
							& $r_V$	& $1.40\pm 0.28$ 	& $\dots$ & $\dots$ & 1.82 & 1.55 & 1.93					
\end{tabular}
\end{ruledtabular}
\end{table}

Regarding the $D_{(s)}\to V$ transition form factors $A_1$, $A_2$, and $V$, it is more interesting to compare their ratios at maximum recoil. The ratios are defined as follows:
\be
r_2 = \frac{A_2(q^2=0)}{A_1(q^2=0)},\qquad r_V = \frac{V(q^2=0)}{A_1(q^2=0)}.
\en
In Table~\ref{tab:form_factors_comparison_DV}, we compare these ratios with the world average given by the PDG~\cite{Tanabashi:2018oca} and with other theoretical results obtained in CQM~\cite{Melikhov:2000yu}, LFQM~\cite{Verma:2011yw}, HM$\chi$T~\cite{Fajfer:2005ug}, and LQCD~\cite{Donald:2013pea}. Our results for the form factor ratios $r_2$ and $r_V$ agree well with the PDG data within uncertainty except for the ratio $r_V(D_s^+ \to \phi)$, for which our prediction is much lower than that from PDG. 
\begin{figure*}[htbp]
\includegraphics[width=0.45\textwidth]{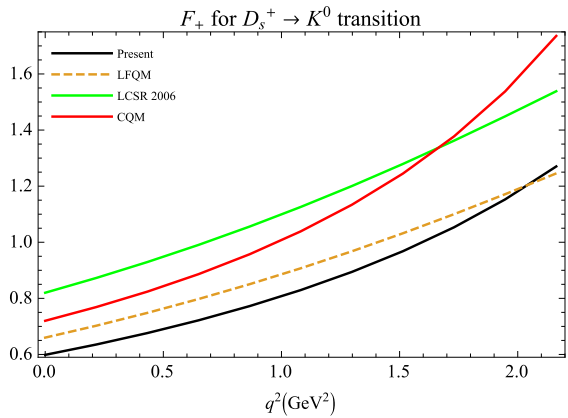}
\vspace*{-5mm}
\caption{Form factor $F_+(q^2)$ for $D_s^+ \to K^0$ in our model, LFQM~\cite{Verma:2011yw}, LCSR~\cite{Wu:2006rd}, and CQM~\cite{Melikhov:2000yu}.}
\label{fig:Ds_K}
\end{figure*}
\begin{figure*}[htbp]
\renewcommand{\arraystretch}{0.3}
\begin{tabular}{cc}
\includegraphics[width=0.45\textwidth]{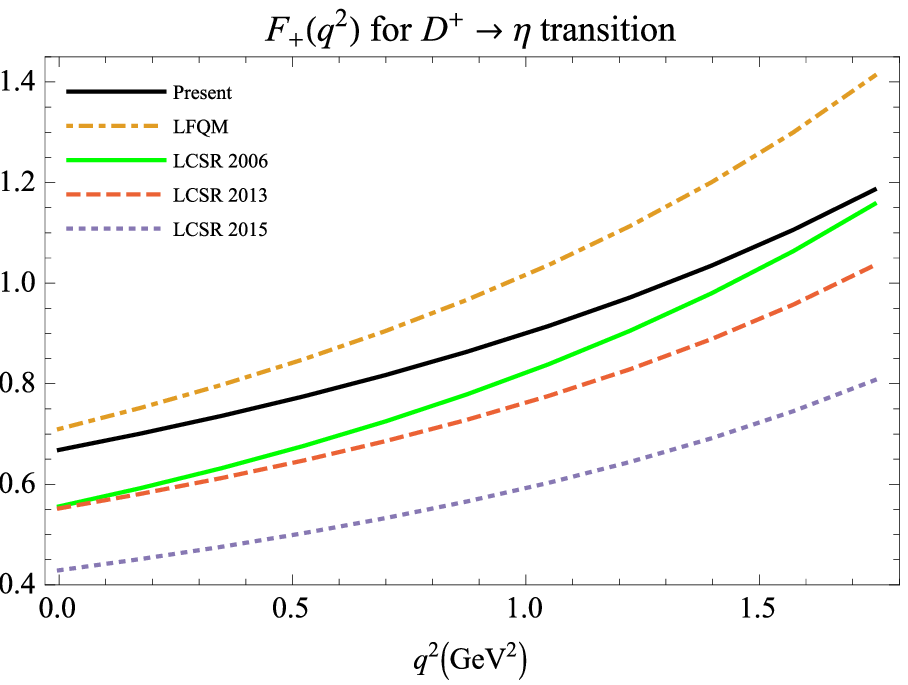}
& \includegraphics[width=0.45\textwidth]{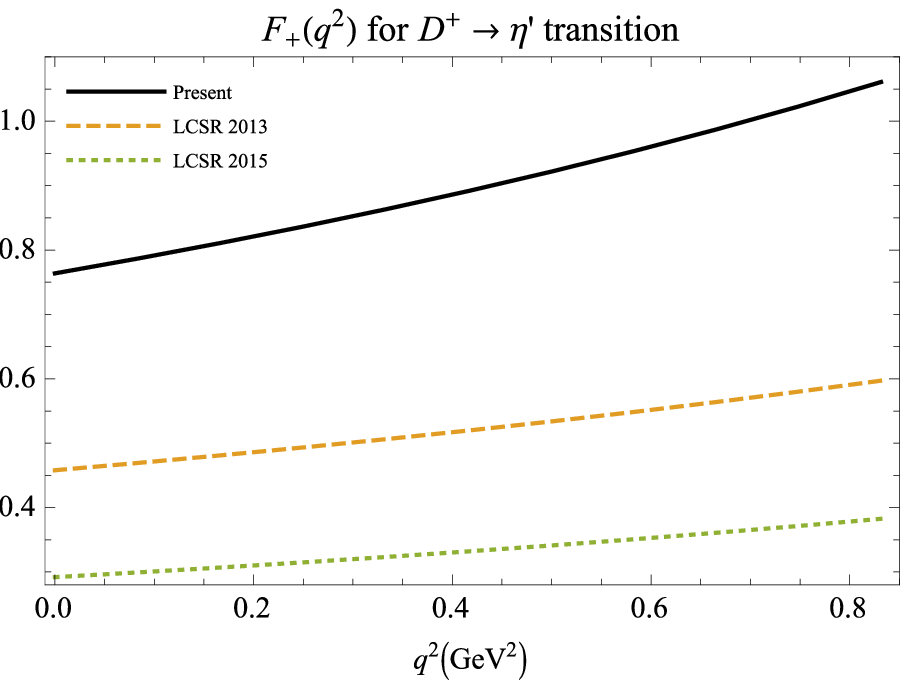}\\
\includegraphics[width=0.45\textwidth]{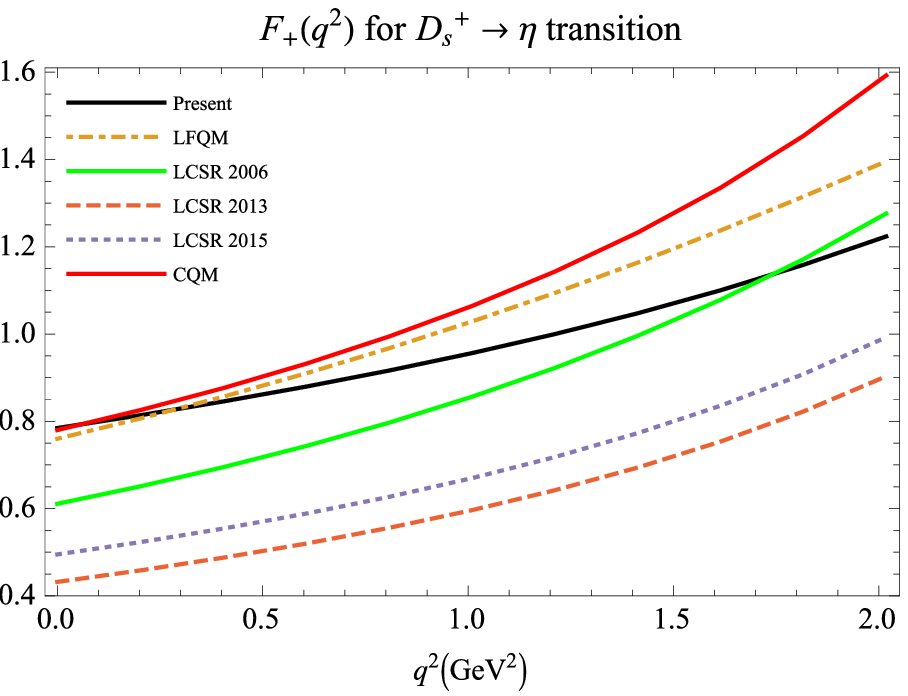}
& \includegraphics[width=0.45\textwidth]{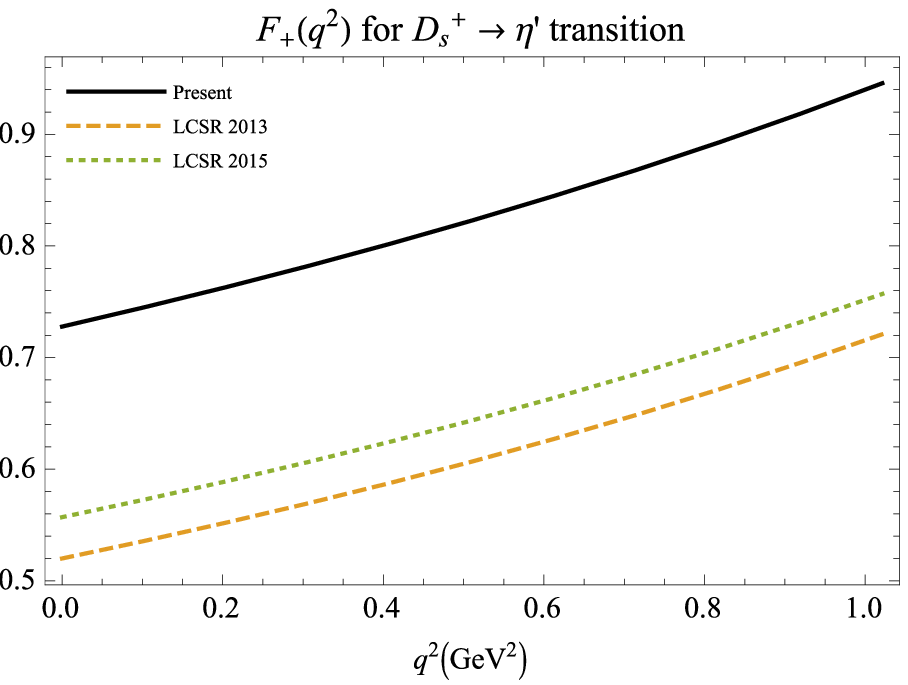}
\end{tabular}
\vspace*{-3mm}
\caption{Form factor $F_+(q^2)$ for $D_{(s)}^+ \to \eta^{(\prime)}$ in our model, LCSR~\cite{Wu:2006rd, Offen:2013nma, Duplancic:2015zna}, and CQM~\cite{Melikhov:2000yu}.}
\label{fig:D_Eta}
\end{figure*}
Note that our prediction $r_V(D_s^+ \to \phi)=1.34$ is close to the value $1.42$ from the LFQM~\cite{Verma:2011yw}. It is also seen that for most cases, the HM$\chi$T predictions~\cite{Fajfer:2005ug} for the ratios at $q^2=0$ are largely different from the PDG values, demonstrating the fact that this model is more suitable for the high $q^2$ region.

\begin{figure*}[htbp]
\renewcommand{\arraystretch}{0.3}
\begin{tabular}{cc}
\includegraphics[width=0.45\textwidth]{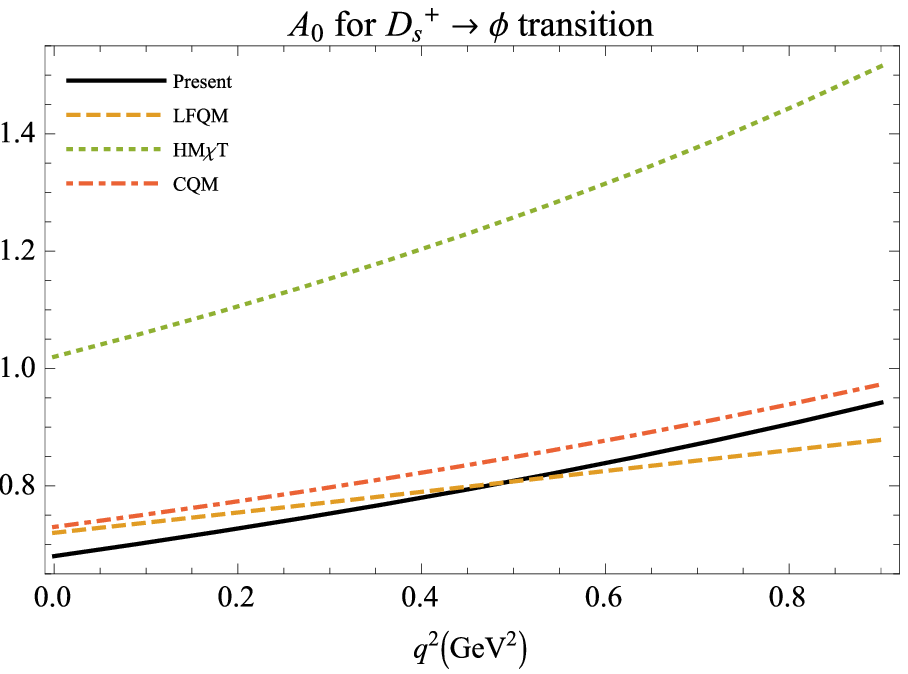}&
\includegraphics[width=0.45\textwidth]{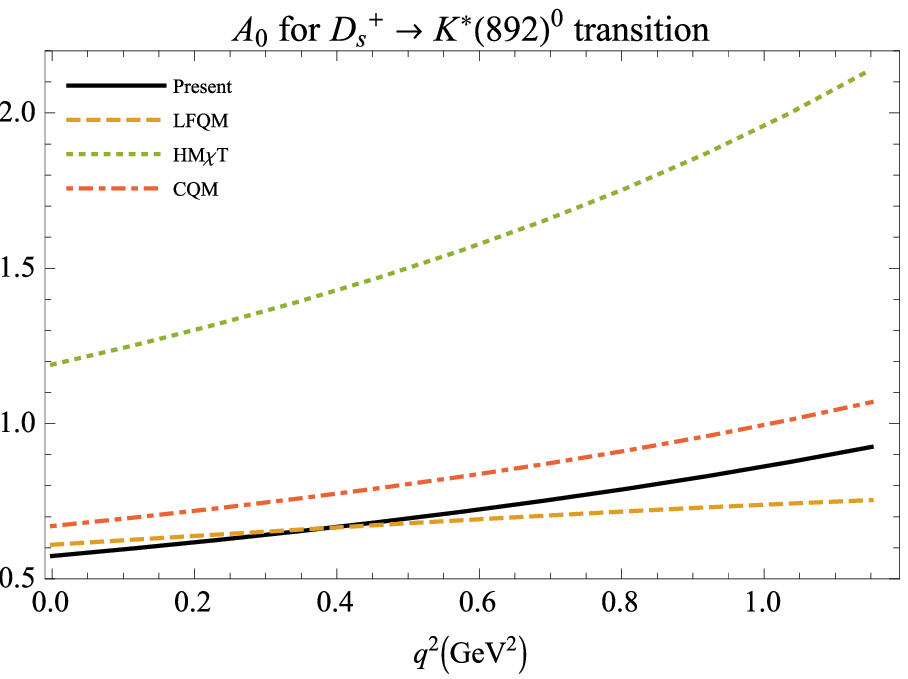}\\
\includegraphics[width=0.45\textwidth]{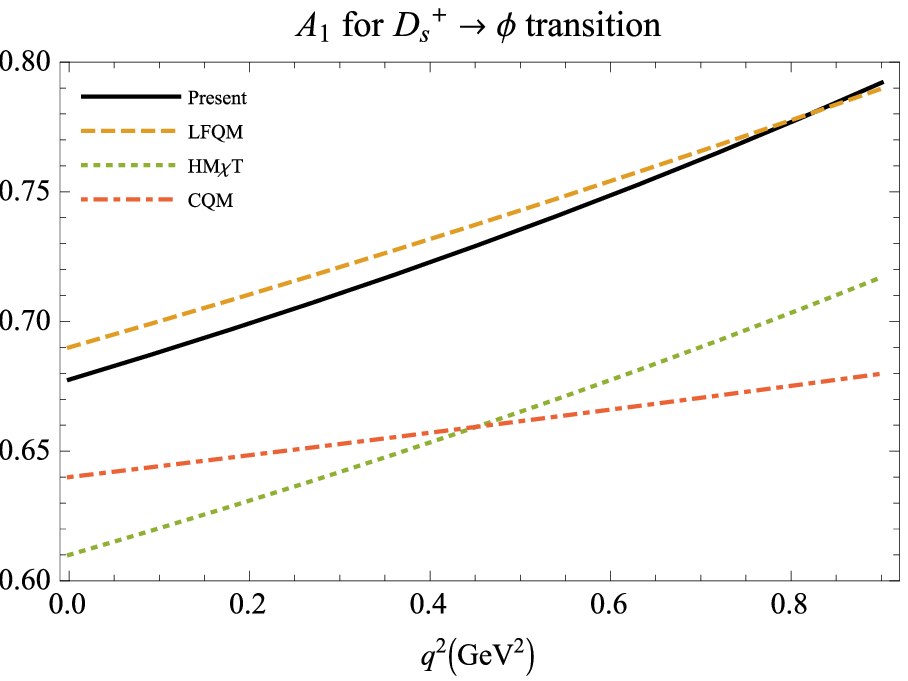}
&\includegraphics[width=0.45\textwidth]{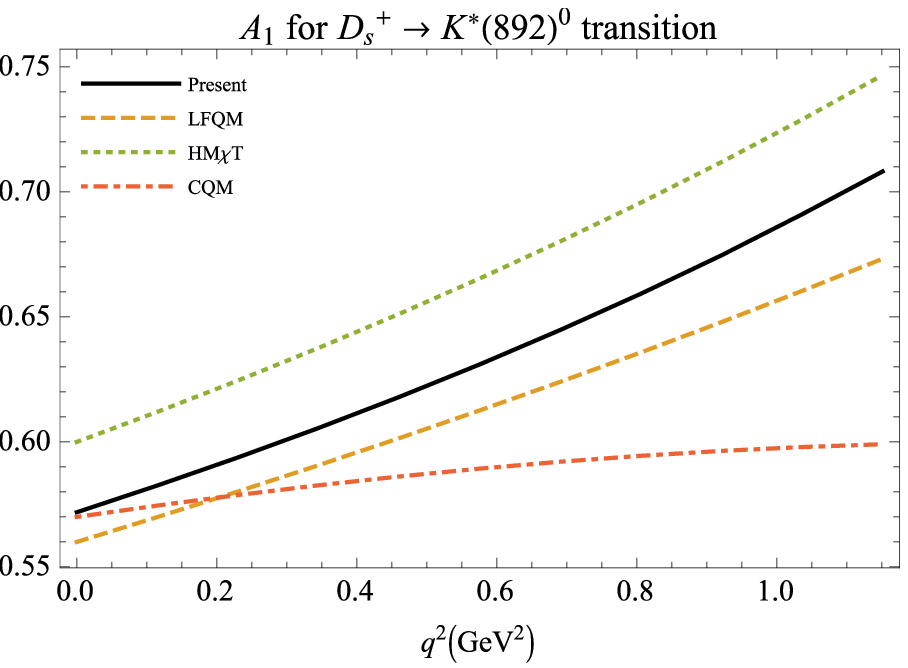}\\
\includegraphics[width=0.45\textwidth]{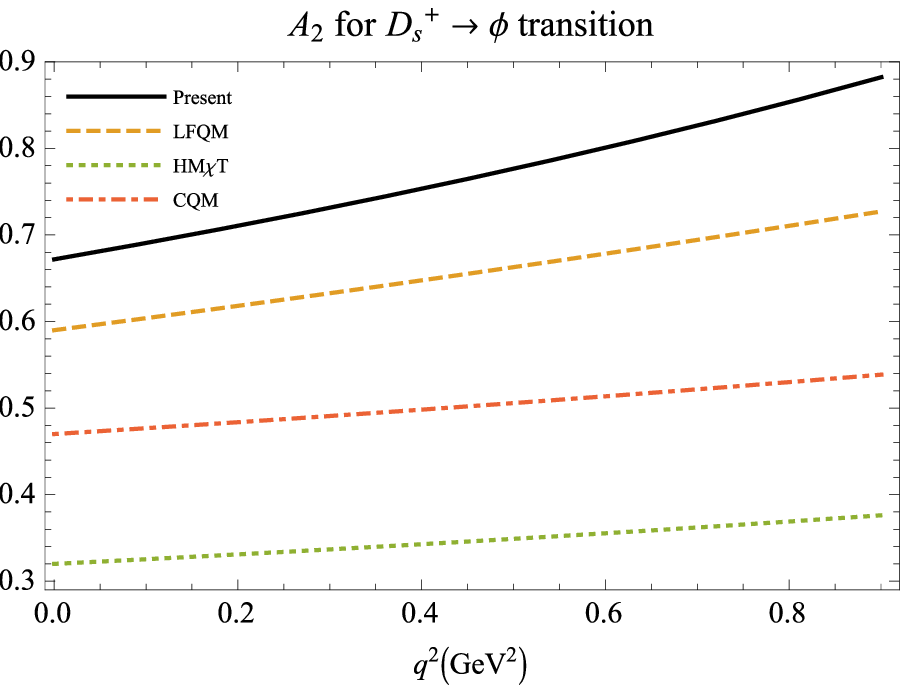}
&\includegraphics[width=0.45\textwidth]{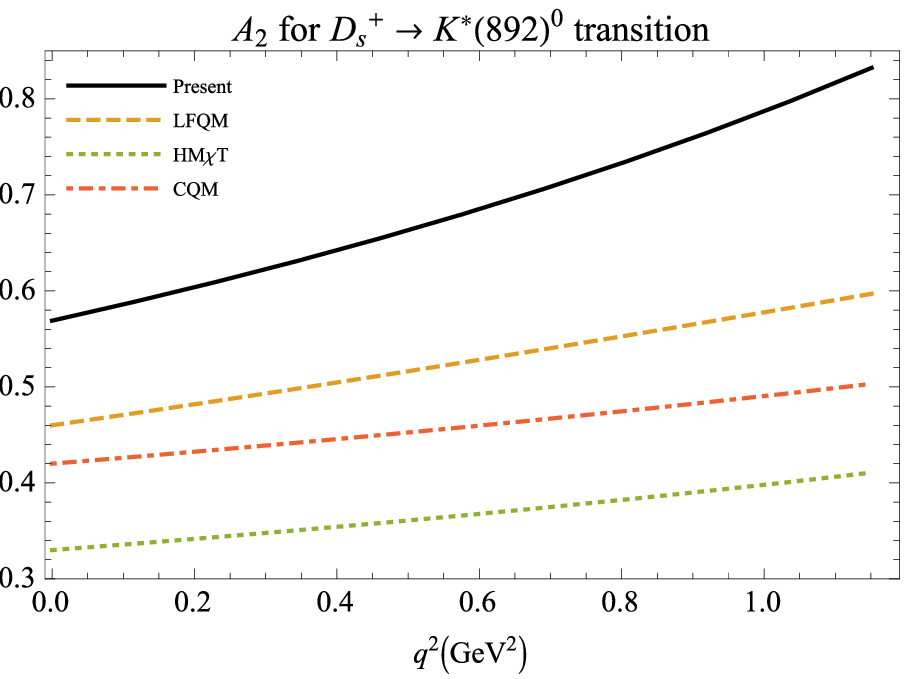}\\
\includegraphics[width=0.45\textwidth]{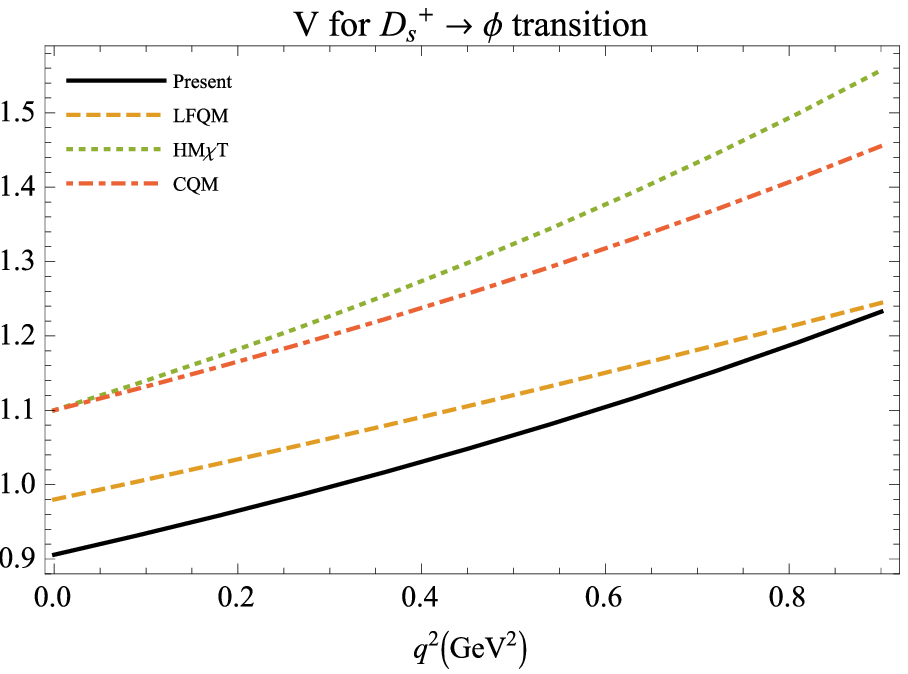}
&\includegraphics[width=0.45\textwidth]{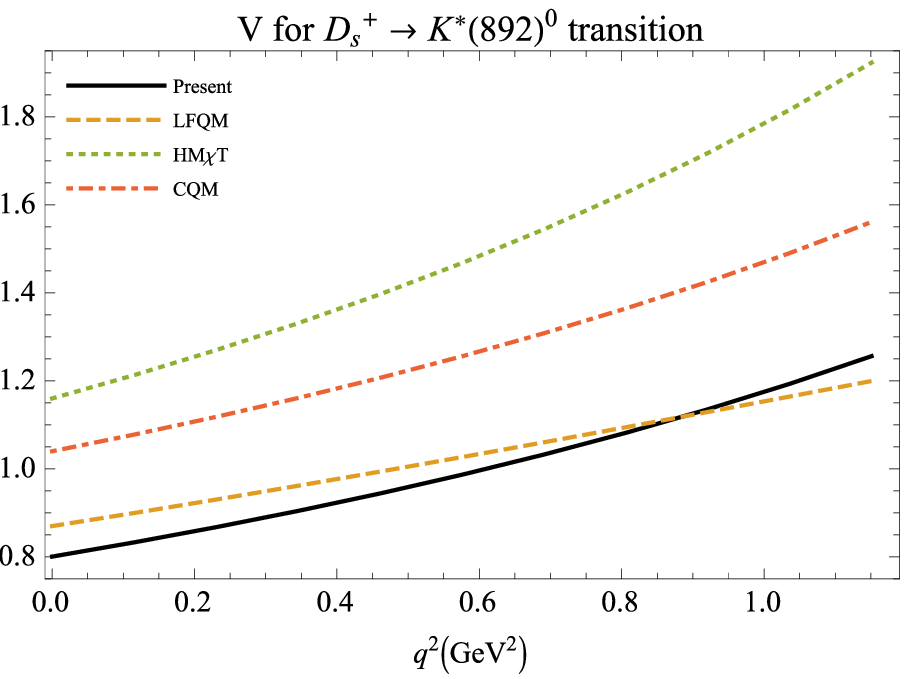}
\end{tabular}
\vspace*{-3mm}
\caption{Form factors for $D_s^+ \to \phi$ (left) and $D_s^+ \to K^*(892)^0$ (right) in our model, LFQM~\cite{Verma:2011yw}, HM$\chi$T~\cite{Fajfer:2005ug}, and CQM~\cite{Melikhov:2000yu}.}
\label{fig:Ds_phi_Kv}
\end{figure*}

\begin{figure*}[htbp]
\renewcommand{\arraystretch}{0.3}
\begin{tabular}{cc}
\includegraphics[width=0.45\textwidth]{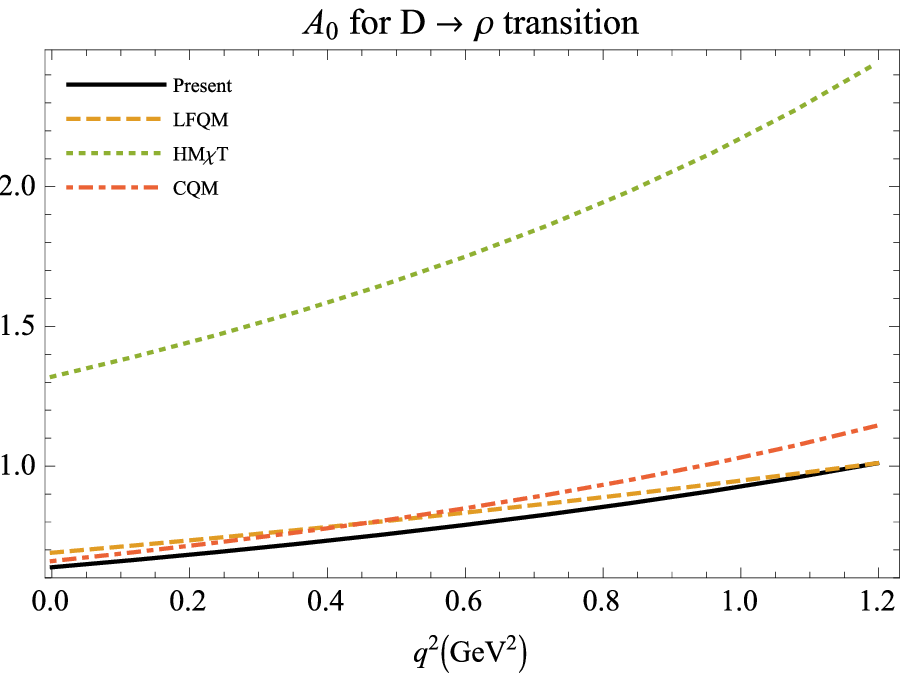}&
\includegraphics[width=0.45\textwidth]{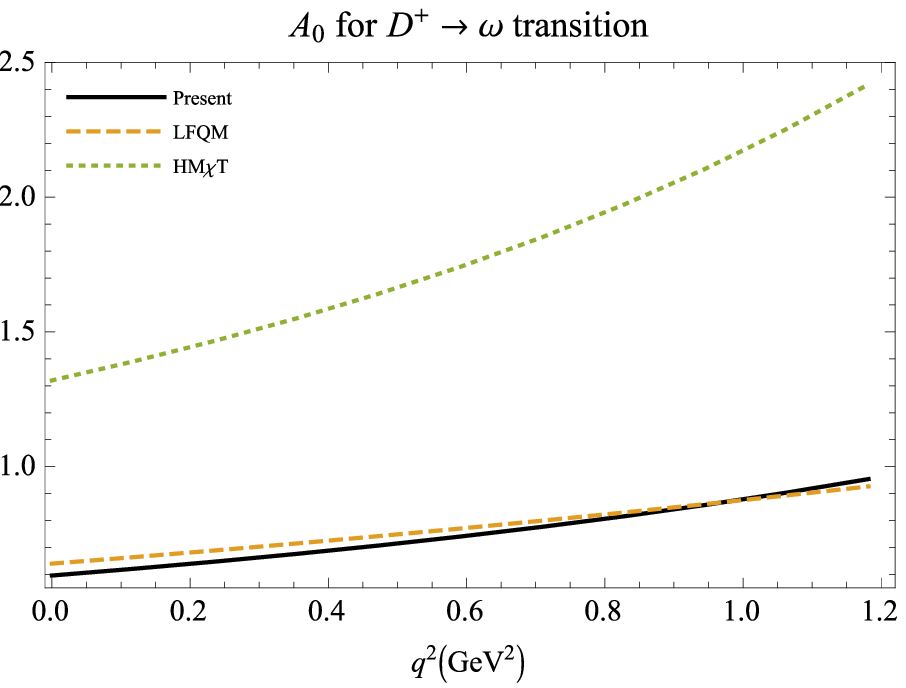}\\
\includegraphics[width=0.45\textwidth]{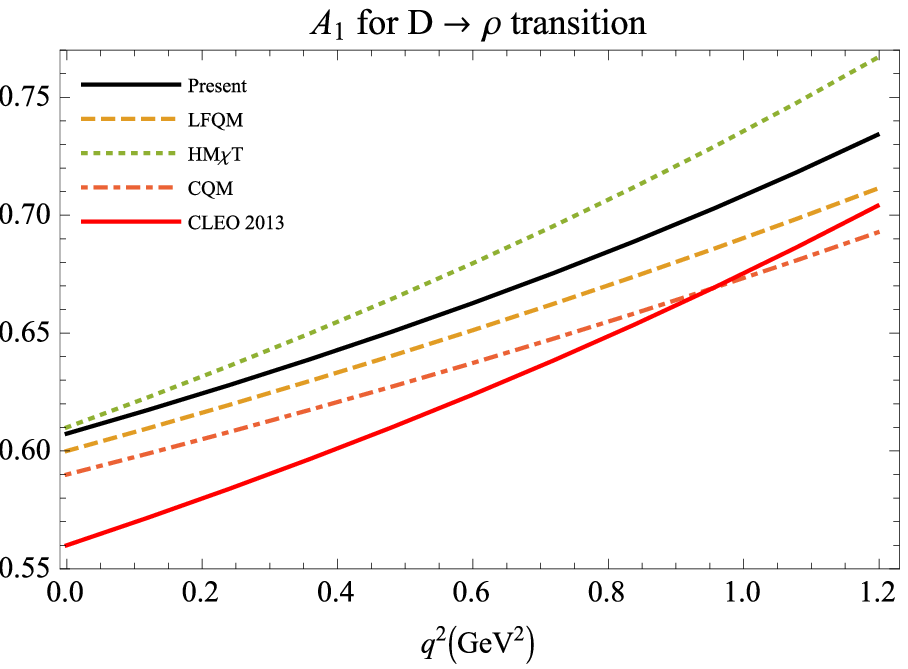}
&\includegraphics[width=0.45\textwidth]{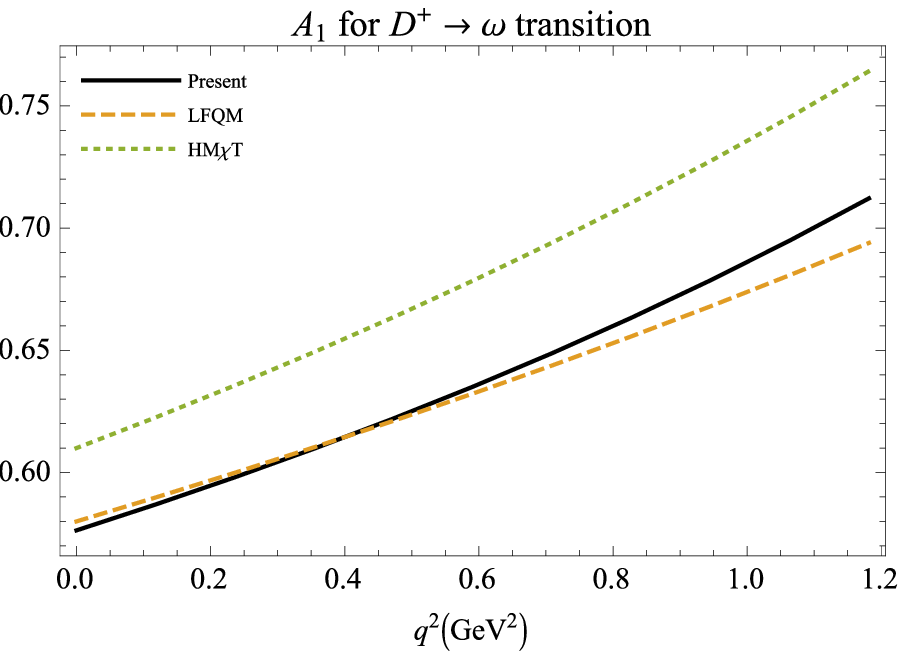}\\
\includegraphics[width=0.45\textwidth]{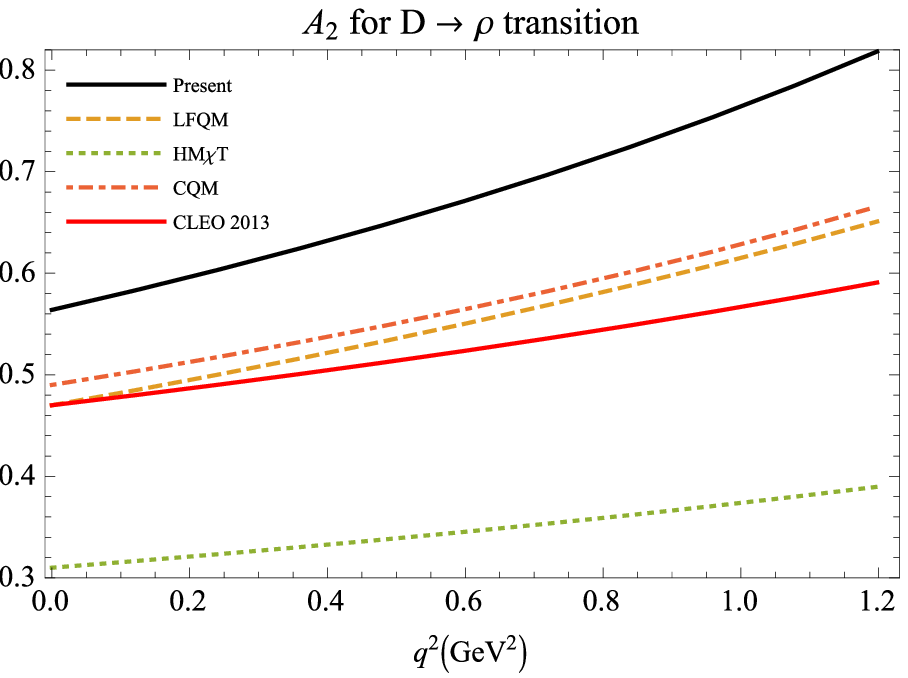}
&\includegraphics[width=0.45\textwidth]{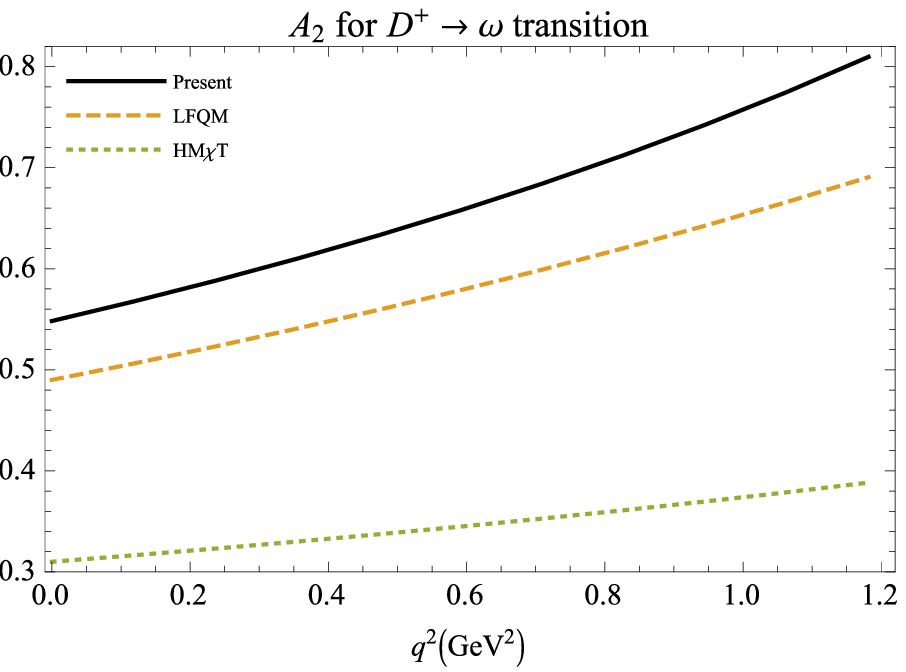}\\
\includegraphics[width=0.45\textwidth]{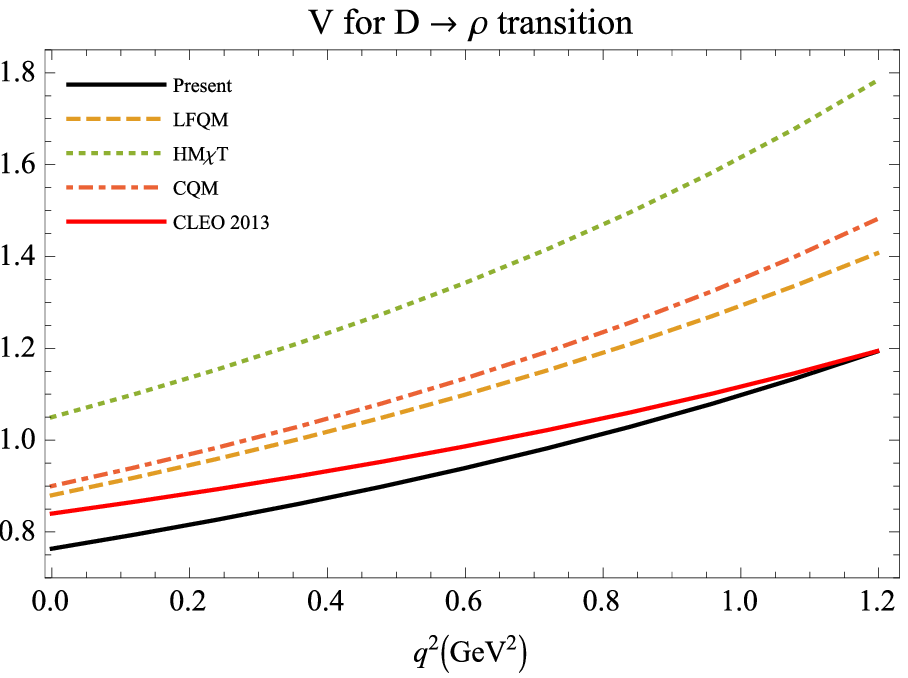}
&\includegraphics[width=0.45\textwidth]{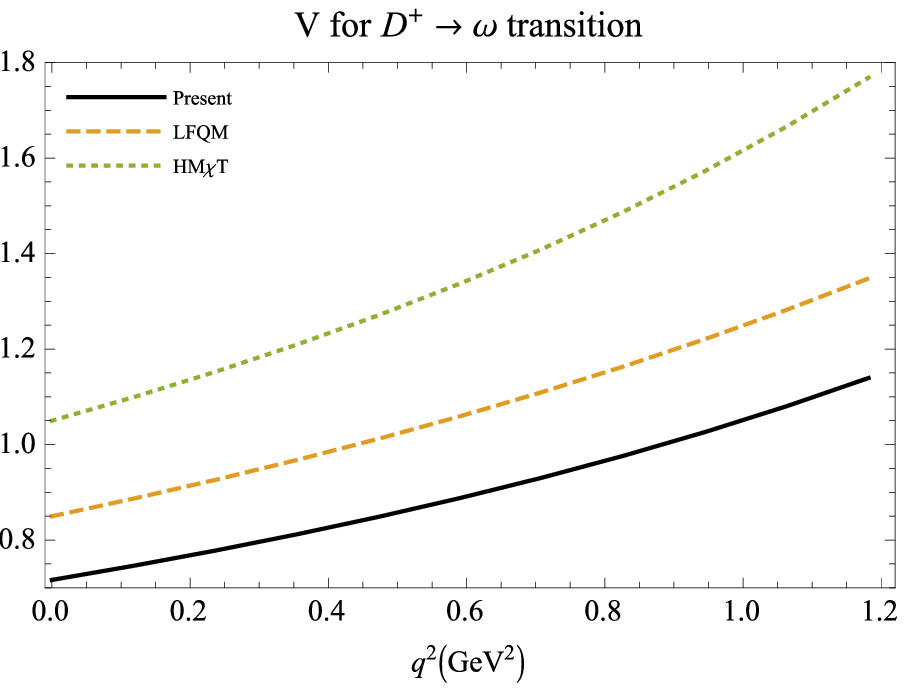}
\end{tabular}
\vspace*{-3mm}
\caption{Form factors for $D \to \rho$ (left) and $D^+ \to \omega$ (right) in our model, LFQM~\cite{Verma:2011yw}, HM$\chi$T~\cite{Fajfer:2005ug}, CQM~\cite{Melikhov:2000yu}, and CLEO data~\cite{CLEO:2011ab}.}
\label{fig:D_rho_omeg}
\end{figure*}

In order to have a better picture of the form factors in the whole $q^2$ range
$0 \leq q^2 \leq q^2_{max} = (m_{D_{(s)}} - m_{P/V})^2$ we plot in Figs.~\ref{fig:Ds_K}--\ref{fig:D_rho_omeg} their $q^2$ dependence  from various studies. It is very interesting to note that, in all cases, our form factors are close to those obtained in the covariant LFQM~\cite{Verma:2011yw}, and this is not for the first time such a good agreement is observed. In a previous study of the semileptonic decays $B_c\to J/\psi(\eta_c)\ell\nu$~\cite{Tran:2018kuv} it was seen that the corresponding form factors agree very well between our model and the covariant LFQM~\cite{Wang:2008xt}. This suggests that a comparison of the two models in more detail may be fruitful. It is also worth noting that the HM$\chi$T~\cite{Fajfer:2005ug} prediction for the form factor $A_0(q^2)$ is systematically much higher than that from other theoretical calculations.



Very recently, the ETM collaboration has provided the lattice determination~\cite{ETM} for the full set of the form factors characterizing the semileptonic $D\to \pi (K)\ell\nu$ and rare $D\to \pi (K)\ell\ell$ decays within and beyond the SM, when an additional tensor coupling is considered. As mentioned before, the decays $D\to \pi (K)\ell\nu$ have been studied in our model already~\cite{Soni:2017eug}. However, we compute the $D\to \pi (K)\ell\nu$ form factors including the tensor one in this paper, in order to compare with the recent ETM results. This demonstrates the fidelity of the CCQM predictions for the hadronic form factors and helps us better estimate the theoretical uncertainties of our model. Moreover, the tensor and scalar form factors are essential for the study of possible new physics in these decays [for more detail we refer to a similar calculation of the full set of $B\to D^{(*)}$ and $B\to \pi ({\rho})$ form factors in our model~\cite{Ivanov:2016qtw, Ivanov:2017hun}].

The new tensor form factor is defined by
\be
\langle P(p_2)|\bar{q}\sigma^{\mu\nu}(1-\gamma^5)c|D(p_1)\rangle 
=\frac{iF^T(q^2)}{M_1+M_2}\left(P^\mu q^\nu - P^\nu q^\mu 
+i \varepsilon^{\mu\nu Pq}\right).
\en
Note that we obtained $F_0(q^2)$ by using the form factors $F_+(q^2)$ and $F_-(q^2)$ defined in Eq.~(\ref{eq:formfac}), with the help of the relation
\be
F_0(q^2) = F_+(q^2) + \frac{q^2}{M_1^2-M_2^2} F_-(q^2).
\en
Meanwhile, the ETM collaboration directly calculated the scalar matrix element $\langle P(p_2)|\bar{q}c|D(p_1)\rangle$ and then determined $F_0(q^2)$ using the equation of motion. In this way, the final result becomes sensitive to the quark mass difference.
\begin{figure*}[htbp]
\renewcommand{\arraystretch}{0.3}
\begin{tabular}{cc}
\includegraphics[width=0.45\textwidth]{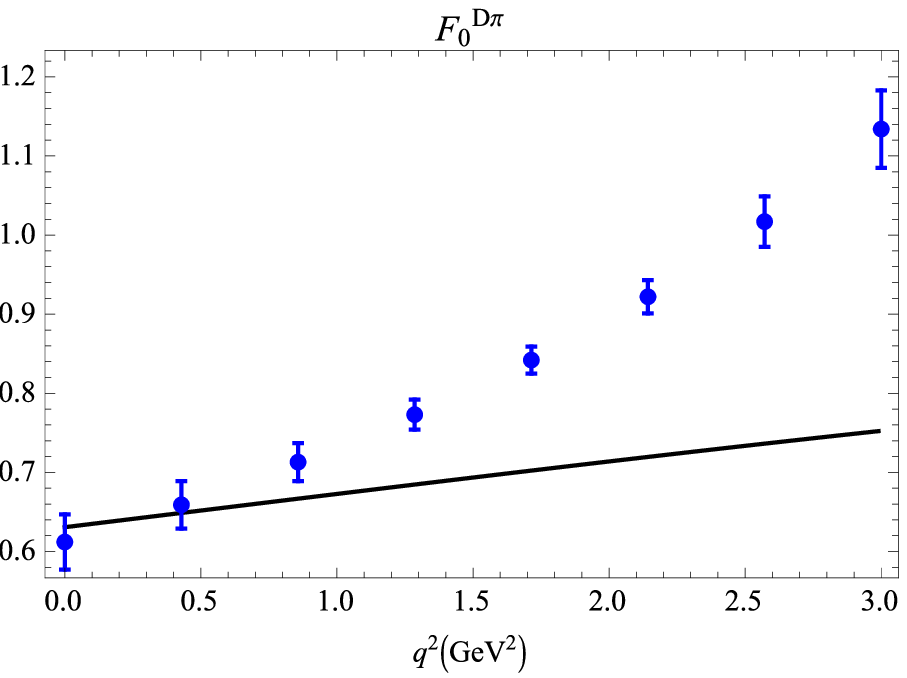}&
\includegraphics[width=0.45\textwidth]{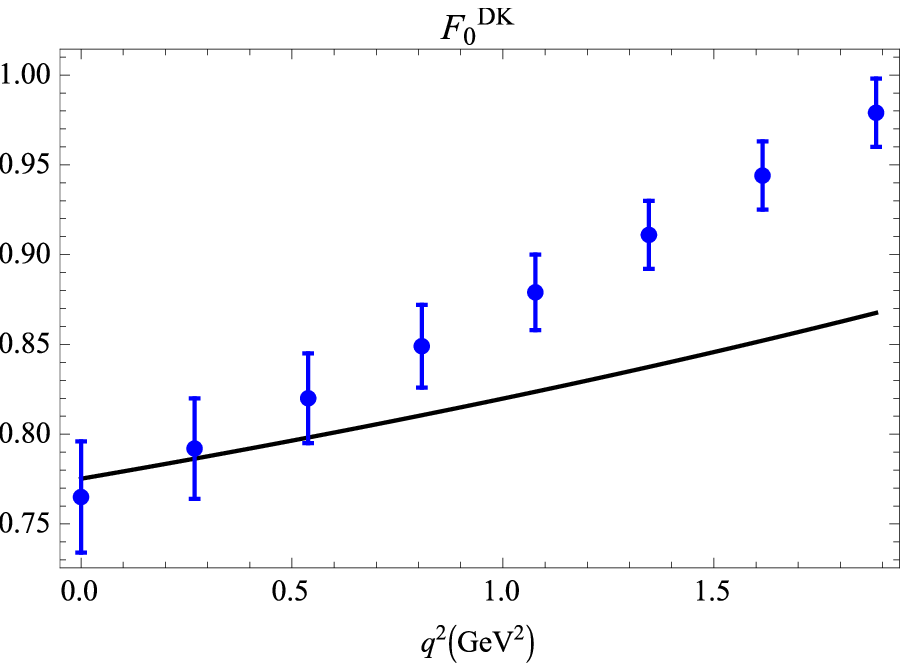}\\
\includegraphics[width=0.45\textwidth]{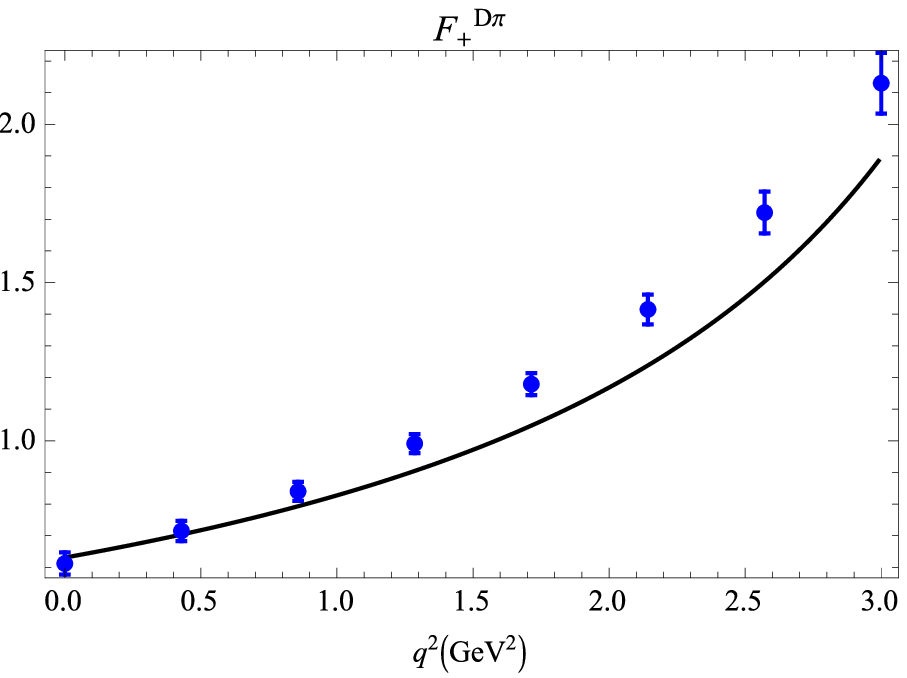}&
\includegraphics[width=0.45\textwidth]{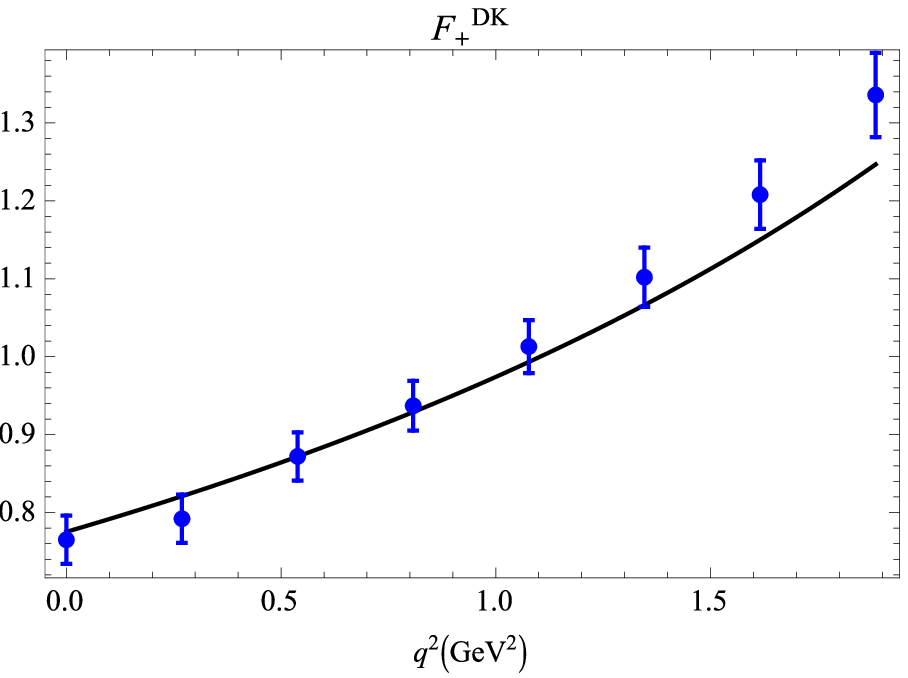}\\\includegraphics[width=0.45\textwidth]{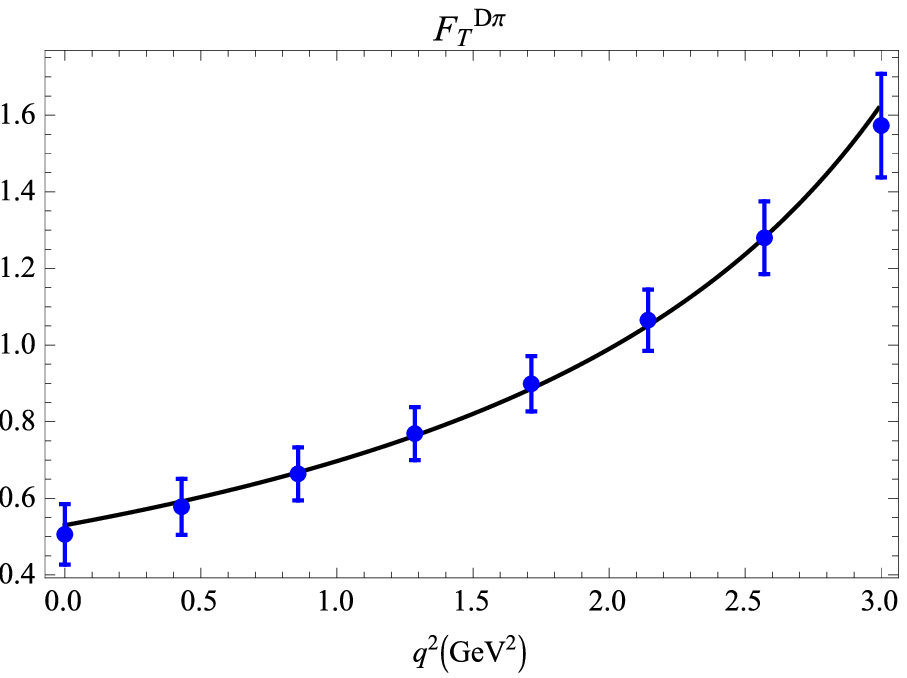}&
\includegraphics[width=0.45\textwidth]{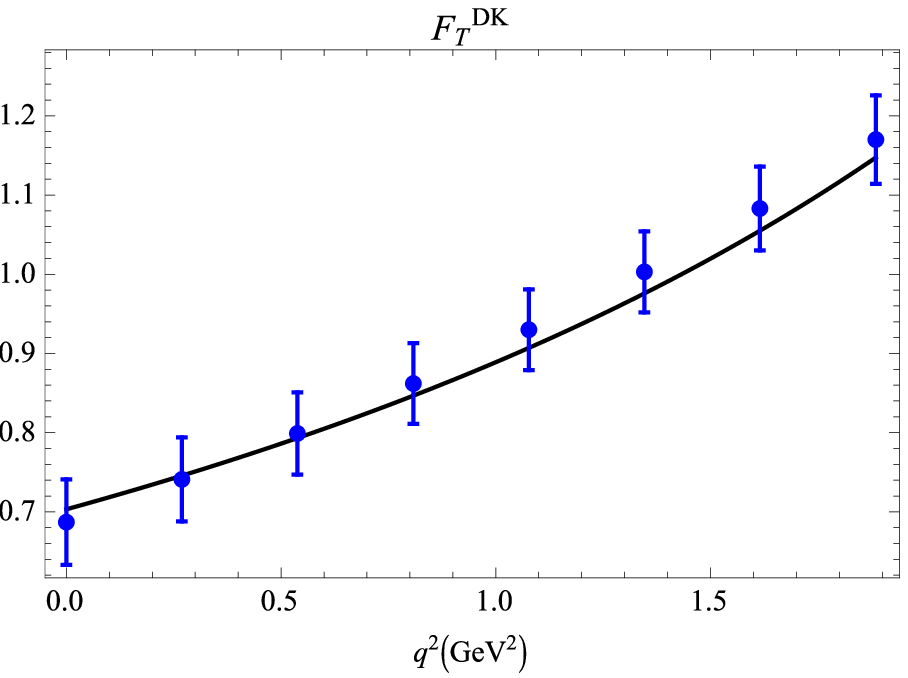}
\end{tabular}
\vspace*{-3mm}
\caption{$D\to \pi(K)\ell\nu$ form factors obtained in our model (solid lines) and in lattice calculation (dots with error bars) by the ETM collaboration~\cite{ETM}.}
\label{fig:D-Pi-K}
\end{figure*}

In Fig.~\ref{fig:D-Pi-K} we compare the form factors $F_0(q^2)$, $F_+(q^2)$, and $F_T(q^2)$ of the $D\to \pi (K)\ell\nu$ transitions with those obtained by the ETM collaboration. It is seen that our $F_0(q^2)$ agrees well with the ETM only in the low $q^2$ region. However, our results for $F_+(q^2)$ are very close to those of the ETM. Note that the determination of $F_+(q^2)$ by the ETM is dependent on $F_0(q^2)$. It is interesting that the tensor form factors between the two studies are in perfect agreement. Even though this form factor does not appear within the SM, this agreement has an important meaning because, in both approaches, the tensor form factor is determined directly from the corresponding matrix element without any additional assumptions. In Table~\ref{tab:scalar}, we present the values of the form factors and their ratios at maximum recoil. One sees that our results agree with the ETM calculation within uncertainty.
\begin{table*}[!htb]
\caption{$D\to \pi(K)\ell\nu$ form factors and their ratios at $q^2=0$.}
\label{tab:scalar}
\renewcommand{\arraystretch}{0.7}
\begin{ruledtabular}
\begin{tabular}{ccccccc}
&$f_+^{D\pi}(0)$ & $f_+^{DK}(0)$ & $f_T^{D\pi}(0)$ & $f_T^{DK}(0)$ &$f_T^{D\pi}(0)/f_+^{D\pi}(0)$ & $f_T^{DK}(0)/f_+^{DK}(0)$\\
\hline
Present & 0.63 & 0.78 & 0.53 & 0.70 & 0.84 & 0.90\\
ETM~\cite{ETM} & $0.612(35)$ & $0.765(31)$ & $0.506(79)$ & $0.687(54)$ & $0.827(114)$ & $0.898(50)$
\end{tabular}
\end{ruledtabular}
\end{table*}

\subsection{Branching fractions and other observables}
\label{subsec:BF}
\begin{table*}[!htb]
\caption{Branching fractions of $D^+(D^0)$-meson semileptonic decays.}
\label{tab:D_branching}
\renewcommand{\arraystretch}{0.7}
\begin{ruledtabular}
\begin{tabular}{lccllll}
Channel 										& Unit 			& Present 	&  Other & Reference 	& Data & Reference\\
\hline
$D^0 \to \rho^- e^+ \nu_e$ 				& $10^{-3}$ 	&   1.62	& 1.97 	& 
$\chi$UA~\cite{Sekihara:2015iha} & $1.445 \pm 0.058 \pm 0.039 $ 		& BESIII~\cite{Ablikim:2018qzz}\\
													& 					& 				& $1.749^{+0.421}_{-0.297} \pm 0.006$& LCSR~\cite{Fu:2018yin} & $1.77 \pm 0.12 \pm 0.10 $ 		& CLEO~\cite{CLEO:2011ab}\\
													& 					& 				& 2.0		& 
													HM$\chi$T~\cite{Fajfer:2005ug}\\
													$D^0 \to \rho^- \mu^+ \nu_\mu$ 		& $10^{-3}$ 	&   1.55 	& 1.84 	& 
$\chi$UA~\cite{Sekihara:2015iha} & & \\
\hline
$D^+ \to \rho^0 e^+ \nu_e$ 			& $10^{-3}$ 	&   2.09 	& 2.54 	& 
$\chi$UA~\cite{Sekihara:2015iha}  &  $1.860 \pm 0.070\pm 0.061$ & 
BESIII~\cite{Ablikim:2018qzz}\\
													& 					& 				& $2.217^{+0.534}_{-0.376} \pm 0.015$& LCSR~\cite{Fu:2018yin} &  $2.17 \pm 0.12^{+0.12}_{-0.22}$ & 
CLEO~\cite{CLEO:2011ab}\\
													& 					& 				& 2.5		& 
													HM$\chi$T~\cite{Fajfer:2005ug}\\
													$D^+ \to \rho^0 \mu^+ \nu_\mu$		& $10^{-3}$ 	&  2.01 	& 2.37 	& 
$\chi$UA~\cite{Sekihara:2015iha}  &  $2.4 \pm 0.4$ 							& 
PDG~\cite{Tanabashi:2018oca}\\
\hline
$D^+ \to \omega e^+ \nu_e$ 			& $10^{-3}$ 	&   1.85 	& 2.46 	& 
$\chi$UA~\cite{Sekihara:2015iha}  &  $1.63 \pm 0.11 \pm 0.08$ 					
& BESIII~\cite{Ablikim:2015gyp}\\
													& 					& 				& 2.5		& 
													HM$\chi$T~\cite{Fajfer:2005ug}	& $1.82 \pm 0.18 \pm 0.07$ 		& CLEO~\cite{CLEO:2011ab}\\
													&				  	&				& 2.1 $\pm$ 0.2 &	
													LFQM~\cite{Cheng:2017pcq}	&  \\
												   $D^+ \to \omega \mu^+ \nu_\mu$	& $10^{-3}$ 	&   1.78 	& 2.29 	& 
$\chi$UA~\cite{Sekihara:2015iha}  &  	& \\
							& 						&					&  2.0 $\pm$ 0.2 &	LFQM~\cite{Cheng:2017pcq}	&  & \\
\hline
$D^+ \to \eta e^+ \nu_e$ 				& $10^{-4}$ 	&  9.37 		& 12 $\pm$ 1	& 
LFQM~\cite{Cheng:2017pcq} & $10.74 \pm 0.81 \pm 0.51$ & BESIII~\cite{Ablikim:2018lfp}\\	
													& 					&  								& $24.5 \pm 5.26$					& LCSR~\cite{Offen:2013nma} & $11.4 \pm 0.9 \pm 0.4$		
													& CLEO~\cite{Yelton:2010js}\\
													&					& & $14.24 \pm 10.98$				& LCSR~\cite{Duplancic:2015zna}\\
$D^+ \to \eta \mu^+ \nu_\mu$ 		& $10^{-4}$ 	&   9.12		& 12 $\pm$ 1 & 
LFQM~\cite{Cheng:2017pcq} & & \\
\hline
$D^+ \to \eta' e^+ \nu_e$ 				& $10^{-4}$ 	&  2.00 		& 1.8 $\pm$ 0.2 & 
LFQM~\cite{Cheng:2017pcq} & $1.91 \pm 0.51 \pm 0.13$	& BESIII~\cite{Ablikim:2018lfp}\\
													& 					& 				&	$3.86 \pm 1.77$					& LCSR~\cite{Offen:2013nma} & $2.16 \pm 0.53 \pm 0.07$	& CLEO~\cite{Yelton:2010js}\\
													&					& & $1.52 \pm 1.17$					& LCSR~\cite{Duplancic:2015zna}\\
$D^+ \to \eta' \mu^+ \nu_\mu$ 		& $10^{-4}$ 	&  1.90		& 	1.7 $\pm$ 0.2 & 
LFQM~\cite{Cheng:2017pcq} &  & 
\end{tabular}
\end{ruledtabular}
\end{table*}
In Tables~\ref{tab:D_branching} and~\ref{tab:Ds_branching}, we summarize our predictions for the semileptonic branching fractions of the $D$ and $D_s$ mesons, respectively. For comparison, we also list results of other theoretical calculations and the most recent experimental data given by the CLEO and BESIII collaborations. Note that the uncertainties of our predictions for the branching fractions and other polarization observables are of order $50\%$, taking into account only the main source of uncertainties related to the form factors.

\begin{table*}[!htb]
\caption{Branching fractions of $D_s$-meson semileptonic decays (in $\%$).}
\label{tab:Ds_branching}
\renewcommand{\arraystretch}{0.7}
\begin{ruledtabular}
\begin{tabular}{lcllll}
Channel & Present & Other & Reference & Data & Reference\\
\hline
$D_s^+ \to \phi e^+ \nu_{e}$			& 3.01		& 2.12 				& $\chi$UA~\cite{Sekihara:2015iha} 	& $2.26 \pm 0.45 \pm 0.09$ 				& BESIII~\cite{Ablikim:2017omq}\\	
													&				& 3.1 $\pm$ 0.3 &	LFQM~\cite{Cheng:2017pcq}	&$2.61 \pm 0.03 \pm 0.08 \pm 0.15$ 	& {\it BABAR}~\cite{Aubert:2008rs}\\
													&				& 	2.4 & HM$\chi$T~\cite{Fajfer:2005ug}		& $2.14 \pm 0.17 \pm 0.08$				& CLEO~\cite{Hietala:2015jqa} \\	
$D_s^+ \to \phi \mu^+ \nu_{\mu}$	&  2.85		& 1.94 & $\chi$UA~\cite{Sekihara:2015iha}	& & \\
													& 				& 2.9 $\pm$ 0.3 &	LFQM~\cite{Cheng:2017pcq}	& $1.94 \pm 0.53 \pm 0.09$ 	&  BESIII~\cite{Ablikim:2017omq} \\
\hline
$D_s^+ \to K^0 e^+ \nu_e$				& 0.20		& 0.27 $\pm$ 0.02 &	
LFQM~\cite{Cheng:2017pcq}	&$0.39 \pm 0.08 \pm 0.03$ 				
& CLEO~\cite{Hietala:2015jqa}\\
$D_s^+ \to K^0  \mu^+ \nu_{\mu}$	& 0.20		& 0.26 $\pm$ 0.02 &	
LFQM~\cite{Cheng:2017pcq}		& \\	
\hline										
$D_s^+ \to K^{*0} e^+ \nu_e$	& 0.18		& 0.202 & $\chi$UA~\cite{Sekihara:2015iha}	& $0.18 \pm 0.04 \pm 0.01$ 				& CLEO~\cite{Hietala:2015jqa}\\
													& 				& 0.19 $\pm$ 0.02 &	LFQM~\cite{Cheng:2017pcq}	\\
													& 				& 0.22 & HM$\chi$T~\cite{Fajfer:2005ug}\\
$D_s^+ \to K^{*0} \mu^+ \nu_{\mu}$		& 0.17	& 0.189 & $\chi$UA~\cite{Sekihara:2015iha} &  &\\
													& 				& 0.19 $\pm$ 0.02 &	LFQM~\cite{Cheng:2017pcq}	\\
\hline
$D_s^+ \to \eta e^+ \nu_e$ 			&  	2.24 	& 2.26 $\pm$ 0.21 & LFQM~\cite{Cheng:2017pcq} & $2.30 \pm 0.31 \pm 0.08$	& BESIII~\cite{Ablikim:2016rqq}\\
													&				& $2.00 \pm 0.32$ 	& LCSR~\cite{Offen:2013nma}& $2.28 \pm 0.14 \pm 0.19$	& CLEO~\cite{Hietala:2015jqa}	\\		
													&				& $2.40 \pm 0.28$					& 
													LCSR~\cite{Duplancic:2015zna}\\
$D_s^+ \to \eta \mu^+ \nu_\mu$ 	&  2.18		& 2.22 $\pm$ 0.20 & LFQM~\cite{Cheng:2017pcq}  & $2.42 \pm 0.46 \pm 0.11$	& BESIII~\cite{Ablikim:2017omq}\\
\hline
$D_s^+ \to \eta' e^+ \nu_e$			&  0.83 		& 0.89 $\pm$ 0.09 & LFQM~\cite{Cheng:2017pcq}  & $0.93 \pm 0.30 \pm 0.05$	& BESIII~\cite{Ablikim:2016rqq}\\
													&				& $0.75 \pm 0.23$					& 
													LCSR~\cite{Offen:2013nma} & $0.68 \pm 0.15 \pm 0.06$	& CLEO~\cite{Hietala:2015jqa}	\\					
													&				& $0.79 \pm 0.14$						& 
													LCSR~\cite{Duplancic:2015zna}\\
$D_s^+ \to \eta' \mu^+ \nu_\mu$ 	&  0.79		& 0.85 $\pm$ 0.08 & LFQM~\cite{Cheng:2017pcq} & $1.06 \pm 0.54 \pm 0.07$	& BESIII~\cite{Ablikim:2017omq}
\end{tabular}
\end{ruledtabular}
\end{table*}
In general, our results for the branching fractions are consistent with experimental data as well as with other theoretical calculations. It is worth mentioning that, for such a large set of decays considered in this study, our branching fractions agree very well with all available experimental data except for one channel, the $D_s^+ \to K^0 \ell^ + \nu_\ell$. In this case, our prediction is nearly twice as small as the CLEO central value~\cite{Hietala:2015jqa} and about $30\%$ smaller than the LFQM prediction~\cite{Cheng:2017pcq}.

We also give prediction for the ratio $\Gamma(D^0 \to \rho^- e^+ \nu_e)/2 \Gamma(D^+ \to \rho^0 e^+ \nu_e)$ which should be equal to unity in the SM, assuming isospin invariance. Our calculation yields $0.98$, in agreement with CLEO's result of $1.03 \pm 0.09^{+0.08}_{-0.02}$~\cite{CLEO:2011ab}. Besides, our ratio of branching fractions $\mathcal B(D_s^+ \to \eta' e^+ \nu_e)/\mathcal B(D_s^+ \to \eta e^+ \nu_e) = 0.37$ coincides with the result $0.36\pm 0.14$ obtained by CLEO~\cite{Yelton:2009aa} and the more recent value $0.40 \pm 0.14$ by BESIII~\cite{Ablikim:2016rqq}. Finally, we predict $\mathcal B(D^+ \to \eta' e^+ \nu_e)/\mathcal B(D^+ \to \eta e^+ \nu_e) = 0.21$, which agrees very well with the values $0.19\pm 0.05$ and $0.18\pm 0.05$ we got from experimental data by CLEO~\cite{Yelton:2010js} and BESIII~\cite{Ablikim:2018lfp}, respectively. It is worth mentioning here that very recently, the BESIII collaboration has reported their measurement of $\mathcal{B}(D^0\to K^-\mu^+\nu_\mu)$~\cite{Ablikim:2018evp} with significantly improved precision. In their paper, they also approved the prediction of our model for the ratio $\mathcal{B}(D^0\to K^-\mu^+\nu_\mu)/\mathcal{B}(D^0\to K^-e^+\nu_e)$ provided in Ref.~\cite{Soni:2017eug}.

\begin{table*}[!htb]
\caption{Semileptonic branching fractions for $D_{(s)}^+ \to D^0 \ell^+ \nu_{\ell}$.}
\label{tab:rare_branching}
\renewcommand{\arraystretch}{0.75}
\begin{ruledtabular}
\begin{tabular}{cclccr}
Channel & Present & Other & Reference & Data & Reference\\
\hline
$D^+ \to D^0 e^+ \nu_e$ 		& $2.23 \times 10^{-13}$ 	& $2.78 \times 10^{-13}$ & \cite{Li:2007kgb} & $<1.0\times 10^{-4}$ 		& BESIII~\cite{Ablikim:2017tdj}\\
											&										& $2.71 \times 10^{-13}$				& \cite{Faller:2015oma}\\
$D_s^+ \to D^0 e^+ \nu_e$ 	& $2.52 \times 10^{-8}$		& $(2.97 \pm 0.03) \times 10^{-8}$	&  \cite{Li:2007kgb} & $\dots$ & $\dots$\\
											&										& $3.34 \times 10^{-8}$					& \cite{Faller:2015oma}
\end{tabular}
\end{ruledtabular}
\end{table*}
In Table~\ref{tab:rare_branching}, we present our results for the semileptonic decays $D_{(s)}^+ \to D^0 e^+\nu_e$, which are rare in the SM due to phase-space suppression. These decays are of particular interest since they are induced by the light quark decay, while the heavy quark acts as the spectator. Besides, the small phase space helps reduce the theoretical errors. The first experimental constraint on the branching fraction $\mathcal{B}(D^+ \to D^0 e^+\nu_e)$ was recently obtained by the BESIII collaboration~\cite{Ablikim:2017tdj}. However, the experimental upper limit is still far above the SM predictions. The branching fractions obtained in our model are comparable with other theoretical calculations using the flavor SU(3) symmetry in the light quark sector~\cite{Li:2007kgb, Faller:2015oma}.

\begin{table}[htbp]
\caption{Forward-backward asymmetry and lepton polarization components.}
\renewcommand{\arraystretch}{0.7}
\begin{ruledtabular}
\begin{tabular}{lcrcccc}
&$\left\langle\mathcal{A}_{FB}^e\right\rangle$ & $\left\langle\mathcal{A}_{FB}^\mu \right\rangle$ & $\left\langle P_L^e \right\rangle$ & $\left\langle P_L^\mu \right\rangle$ & $\left\langle P_T^e \right\rangle$ & $\left\langle P_T^\mu \right\rangle$\\\hline
$D^0\to \rho^-\ell^+\nu_\ell$ & $0.21$ & $0.19$ & $-1.00$ & $-0.92$ & $1.4\times 10^{-3}$ & 0.22\\
$D^+\to\rho^0\ell^+\nu_\ell$ & $0.22$ & $0.19$ & $-1.00$ & $-0.92$ & $1.4\times 10^{-3}$ & 0.22\\
$D^+\to\omega\ell^+\nu_\ell$ & $0.21$ & $0.19$ & $-1.00$ & $-0.92$ & $1.4\times 10^{-3}$ & 0.22\\
$D^+\to\eta\ell^+\nu_\ell$ & $-6.4\times 10^{-6}$ & $-0.06$ & $-1.00$ & $-0.83$ & $2.8\times 10^{-3}$ & 0.44\\
$D^+\to\eta^\prime\ell^+\nu_\ell$ & $-13.0\times 10^{-6}$ & $-0.10$ & $-1.00$ & $-0.70$ & $4.2\times 10^{-3}$ & 0.59\\
$D^+\to D^0\ell^+\nu_\ell$ & $-0.10$ & $\dots$ & $-0.72$ & $\dots$ & 0.56 & $\dots$\\
$D_s^+\to \phi\ell^+\nu_\ell$ & $0.18$ & $0.15$ & $-1.00$ & $-0.91$ & $1.5\times 10^{-3}$ & 0.23\\
$D_s^+\to K^{\ast 0}\ell^+\nu_\ell$ & $0.22$ & $0.20$ & $-1.00$ & $-0.92$ & $1.4\times 10^{-3}$ & 0.22\\
$D_s^+\to K^0\ell^+\nu_\ell$ & $-5.0\times 10^{-6}$ & $-0.05$ & $-1.00$ & $-0.86$ & $2.4\times 10^{-3}$ & 0.39\\
$D_s^+\to \eta\ell^+\nu_\ell$ & $-6.0\times 10^{-6}$ & $-0.06$ & $-1.00$ & $-0.84$ & $2.7\times 10^{-3}$ & 0.42\\
$D_s^+\to \eta^\prime\ell^+\nu_\ell$ & $-11.2\times 10^{-6}$ & $-0.09$ & $-1.00$ & $-0.75$ & $3.8\times 10^{-3}$ & 0.54\\
$D_s^+\to D^0\ell^+\nu_\ell$ & $-7.37\times 10^{-4}$ & $\dots$ & $-1.00$ & $\dots$ & 0.038 & $\dots$\\
\end{tabular}
\end{ruledtabular}
\label{tab:pol}
\end{table}
Finally, in Table~\ref{tab:pol} we list our predictions for the forward-backward asymmetry $\langle\mathcal{A}_{FB}^\ell\rangle$, the longitudinal polarization $\langle P_L^\ell \rangle$, and the transverse polarization $\langle P_T^\ell \rangle$ of the charged lepton in the final state. It is seen that, for the $P\to V$ transitions, the lepton-mass effect in $\langle\mathcal{A}_{FB}^\ell\rangle$ is small, resulting in a difference of only $10\%$--$15\%$ between the corresponding electron and muon modes. For the $P\to P^\prime$ transitions, $\langle\mathcal{A}_{FB}^\mu\rangle$ are about $10^4$ times larger than $\langle\mathcal{A}_{FB}^e\rangle$. This is readily seen from Eq.~(\ref{eq:FB}): for $P\to P^\prime$ transitions the two helicity amplitudes $H_{\pm}$ vanish and the forward-backward asymmetry is proportional to the lepton mass squared.  Regarding the longitudinal polarization, the difference between $\langle P_L^\mu \rangle$ and $\langle P_L^e \rangle$ is $10\%$--$30\%$. One sees that the lepton-mass effect in the transverse polarization is much more significant than that in the longitudinal one. This is true for both $P\to P^\prime$ and $P\to V$ transitions. Note that the values of $\langle\mathcal{A}_{FB}^e\rangle$ and $\langle P_{L(T)}^e \rangle$ for the rare decays $D^+_{(s)} \to D^0 e^+\nu_e$ are quite different in comparison with other $P\to P^\prime$ transitions due to their extremely small kinematical regions.
\section{Summary and Conclusion}
\label{sec:conclusion}
We have presented a systematic study of the $D$ and $D_s$ semileptonic decays within the framework of the CCQM. All the relevant form factors are calculated in the entire range of momentum transfer squared. We have also provided a detailed comparison of the form factors with other theoretical predictions and, in some cases, with available experimental data. In particular, we have observed a good agreement with the form factors obtained in the covariant LFQM, for all decays. It is worth noting that our tensor form factors for the $D\to \pi (K)\ell\nu$ decays are in perfect agreement with the recent LQCD calculation by the ETM collaboration~\cite{ETM}. 

We have given our predictions for the semileptonic branching fractions and their ratios. In general, our results are in good agreement with other theoretical approaches and with recent experimental data obtained by {\it BABAR}, CLEO, and BESIII. In all cases, our predictions for the branching fractions agree with experimental data within 10\%, except for the $D_s^+ \to K^0 \ell^ + \nu_\ell$ channel. Our predictions for the ratios of branching fractions are in full agreement with experimental data. To conclude, we have provided the first ever theoretical predictions for the forward-backward asymmetries, and lepton longitudinal and transverse polarizations, which are important for future experiments.

\begin{acknowledgments}
J.~N.~P. acknowledges financial support from University Grants Commission of India under Major Research Project F.No.42-775/2013(SR). P.~S. acknowledges support from Istituto Nazionale di Fisica Nucleare, I.S. QFT\_\,HEP. M.~A.~I., J.~G.~K., and C.~T.~T. thank Heisenberg-Landau Grant for providing support for their collaboration. M.~A.~I. acknowledges financial support of PRISMA Cluster of Excellence at University of Mainz. N.~R.~S. thanks Bogoliubov Laboratory of Theoretical Physics, Joint Institute for Nuclear Research for warm hospitality during Helmholtz-DIAS International Summer School “Quantum Field Theory at the Limits: from Strong Field to Heavy Quarks” where this work was initiated. C.~T.~T. acknowledges support from Duy Tan University during the beginning stage of this work. M.~A.~I. and C.~T.~T. appreciate warm hospitality of Mainz Institute for Theoretical Physics at University of Mainz, where part of this work was done.
\end{acknowledgments}
{\it Note added.}---Recently, we became aware of the paper~\cite{Ablikim:2018upe} where the BESIII collaboration reported their new measurements of the branching fractions for the decays $D_s^+\to K^0 e^+\nu_e$ and $D_s^+\to K^{\ast 0} e^+\nu_e$ with improved precision. They also obtained for the first time the values of the form factors at maximum recoil. Our predictions for the branching fraction $\mathcal{B}(D_s^+\to K^{\ast 0} e^+\nu_e)$ as well as the form factor parameters $f_+^{D_sK}(0)$, $r_V^{D_sK^\ast}(0)$, and $r_2^{D_sK^\ast}(0)$ agree with the new BESIII results. Regarding their result $\mathcal{B}(D_s^+\to K^0 e^+\nu_e)=(3.25\pm 0.41)\times 10^{-3}$, the central value is closer to our prediction, in comparison with the CLEO result~\cite{Hietala:2015jqa}. However, the BESIII result is still at $1\sigma$ larger than ours. 

\end{document}